\documentclass[aps,prd,onecolumn,groupedaddress,nofootinbib,superscriptaddress]{revtex4}  
\usepackage{graphicx,mathtools,tabu}
\usepackage{epstopdf}
\usepackage{amsmath}
\usepackage{amsfonts}
\usepackage[colorlinks=true,citecolor=red,linkcolor=blue,breaklinks=true]{hyperref}
\usepackage{accents}
\usepackage[figure, table]{hypcap}
\usepackage{amssymb}
\usepackage{appendix}
\usepackage{physics}
\usepackage{comment}
\usepackage{bbold}
\usepackage{color}
\usepackage{slashed}
\usepackage{subfigure}
\usepackage{setspace}
\usepackage{footnote}
\usepackage{multirow}
\usepackage[normalem]{ulem}
\usepackage[utf8]{inputenc}
\usepackage{mathrsfs}
\usepackage{longtable}
\usepackage{enumitem}

\LTcapwidth=\textwidth

\usepackage{url}
\usepackage{wrapfig}

 \def\be   {\begin{equation}}  
 \def\ee   {\end{equation}}
 \def\ba   {\begin{array}}     
  \def\ea   {\end{array}}
 \def\bea  {\begin{eqnarray}}  
  \def\eea  {\end{eqnarray}}
 \def\bean {\begin{eqnarray*}}  
 \def\eean {\end{eqnarray*}}

  \def\be {\beta}

\def\to {\rightarrow}

\def \A{\mathcal{A}}
\def \B{\mathcal{B}}

\begin{document}

\title{Visible Neutrino Decays and the Impact of the Daughter-Neutrino Mass}

\author{Andr\'{e} de Gouv\^{e}a}
\email{degouvea@northwestern.edu}
\affiliation{Northwestern University, Department of Physics \& Astronomy, 2145 Sheridan Road, Evanston, IL 60208, USA}
\author{Manibrata Sen}
\email{manibrata@mpi-hd.mpg.de}
\affiliation{Max-Planck-Institut f{\"u}r Kernphysik, Saupfercheckweg 1, 69117 Heidelberg, Germany}
\author{Jean Weill}
\email{jeanweill@u.northwestern.edu}
\affiliation{Northwestern University, Department of Physics \& Astronomy, 2145 Sheridan Road, Evanston, IL 60208, USA}

\begin{abstract}
   We compute the differential decay width of two- and three-body neutrino decays, assuming neutrinos are Dirac fermions and allowing for the possibility that the decay-daughters have nonzero masses. We examine different hypotheses for the interaction that mediates neutrino decay and concentrate on identifying circumstances where the decay-daughters can significantly impact the neutrino-decay signature at different experiments. We are especially interested in decay daughters produced by right-chiral neutrino fields, when the mass of the daughter plays a decisive role. As a concrete example, we compare the effects of visible and invisible antineutrino decays at the JUNO experimental setup.
\end{abstract}

\maketitle

\section{Introduction}
\setcounter{equation}{0}

The discovery of nonzero neutrino masses unlocked several basic questions about neutrino properties, among them whether neutrinos have a finite lifetime. Given everything we know about neutrinos, the answer is, technically, yes. There are at least three neutrino species and at least two of them have nonzero masses~\cite{ParticleDataGroup:2020ssz}. All of these participate in the charged- and neutral-current weak interactions in such a way that, keeping in mind it is established that lepton-flavor numbers are violated by the weak interactions, the two heaviest neutrinos decay into a lighter neutrino and a photon~\cite{Pal:1981rm}. Furthermore, assuming the three neutrinos are not almost degenerate in mass, the three-body decay of one heavier neutrino into three lighter ones is also mediated by weak interactions. The weak-interactions-mediated lifetimes, however, are astronomical, exceeding the age of the universe by many orders of magnitude. There are many reasons for this, including the very small neutrino masses, the weakness of the  weak interactions at the neutrino-mass scale ($M_{W,Z}^2\gg 1$~eV$^2$), and the fact that neutrino decays are flavor-changing neutral currents and hence impacted by the GIM mechanism.

New neutrino interactions can lead to significantly shorter neutrino lifetimes, also for several reasons~\cite{Bahcall:1972my,Schechter:1981cv,Bahcall:1986gq,Nussinov:1987pc,Frieman:1987as,Kim:1990km,Biller:1998nc}. For example, virtually massless new particles provide new neutrino decay modes and the exchange of new quanta much lighter than $W$-bosons and $Z$-bosons can mediate faster decays that are not GIM suppressed~\cite{PhysRevLett.45.1926,GELMINI1981411,Gelmini:1983ea,BERTOLINI1988714,SANTAMARIA1987423}. We discuss these in more detail in Section~\ref{sec:models}. Some are best constrained by searches for new interactions of neutrinos and charged leptons while others are ultimately best constrained by searches for a finite neutrino lifetime. Here, we will be especially interested in identifying and investigating the latter.

Experimentally, there is no evidence for finite neutrino lifetimes. A finite neutrino lifetime will impact direct and indirect measurements of neutrinos that propagate long distances~\cite{1984MNRAS.211..277D, Doroshkevich:1989bf, Berezhiani1992, Fogli:1999qt,Choubey:2000an,Lindner:2001fx,Beacom:2002cb,Joshipura:2002fb,Bandyopadhyay:2002qg, Beacom:2002vi,Beacom:2004yd,Berryman:2014qha,Picoreti:2015ika,FRIEMAN1988115,Mirizzi:2007jd,GonzalezGarcia:2008ru,Maltoni:2008jr,Baerwald:2012kc,Broggini:2012df,Dorame:2013lka,Gomes:2014yua,Abrahao:2015rba,Coloma:2017zpg,Gago:2017zzy,Choubey:2018cfz,deSalas:2018kri, deGouvea:2019goq,Funcke:2019grs,Escudero:2020ped,Abdullahi:2020rge,Akita:2021hqn,Picoreti:2021yct,Chen:2022idm}. These include the relic ``big bang'' neutrinos, neutrinos produced in far-away astrophysical sources, including the Sun, supernova explosions, active galactic nuclei, and man-made neutrinos, as long as these are detected some large distance away from their sources. Roughly speaking, the effects of neutrino decay are of two types: (i) the decay-parents ``disappear'' and (ii) the decay-daughters ``appear.'' The latter case is relevant for phenomenology if the daughters are capable of interacting with the detector of interest, otherwise one need only worry about the disappearing parents.  We discuss how neutrino decay impacts neutrino-oscillation-like experiments in more detail in Section~\ref{sec:formalism}.

We are especially interested in the neutrino-daughters of the decay and whether they can interact with the detectors of interest -- visible daughters -- or not -- invisible daughters.
The nature of the neutrino-daughter depends on the interaction responsible for the neutrino decay and the mass of the daughter relative to that of the parent. We address this in Section~\ref{sec:2body}, for the case of two-body neutrino decays, and Section~\ref{sec:3body}, for the three-body decay. We concentrate on the hypothesis that neutrinos are Dirac fermions. In between, in Section~\ref{sec:JUNO}, we quantify the importance of visible daughters by discussing the sensitivity of JUNO to a finite neutrino lifetime.

Our results are summarized in Section~\ref{sec:conclusion}.

\section{Neutrino-Decay Models}
\label{sec:models}
\setcounter{equation}{0}

We will consider both two-body neutrino decays into a lighter neutrino and a massless scalar $\varphi$ and three-body neutrino decays into lighter neutrinos. 

\subsection{Two-Body-Decay Interactions}

We assume the massless scalar field $\varphi$ to be uncharged under the Standard Model gauge group and ignore all other potential interactions it may have with Standard Model particles, including renormalizable ones like $(H^{\dagger}H)\varphi$ and $(H^{\dagger}H)\varphi^2$, where $H$ is the Standard Model Higgs-boson doublet. We also assume $\varphi$ does not have a nontrivial vacuum expectation value. For all practical purposes, the quanta associated to the $\varphi$ field are completely invisible. 

If the neutrinos are Majorana fermions, assuming there are no other new particles, the interaction between two neutrinos and $\varphi$ is included in the dimension-six (generation indices implied)
\begin{equation}
{\cal L}_{\rm 2b-decay} \propto (HL)(HL)\varphi + h.c.,
\label{eq:maj_su2}
\end{equation} 
where $L$ is a left-chiral lepton doublet, expressed as a two-component Weyl fermion.  After electroweak symmetry breaking, 
\begin{equation}
{\cal L}_{\rm 2b-decay} \propto \nu\nu\varphi + h.c..
\label{eq:maj}
\end{equation} 
We are interested in computing the decay of neutrinos in the laboratory frame, where the neutrinos are ultrarelativistic. In this case, it is meaningful to refer to the left-helicity neutrino as a `neutrino' and the right-helicity neutrino as an `antineutrino.'\footnote{When the left-helicity state interacts with a target via the charged-current weak interactions, it will produce a negatively charged lepton with virtually 100\% probability. The right-helicity state, instead, will mediate the production of positively charged leptons with virtually 100\% probability, exactly like Dirac neutrinos and antineutrinos, respectively. } Using this language, Eq.~(\ref{eq:maj}) mediates the decay of a parent-neutrino into both a daughter-neutrino or a daughter-antineutrino. Here, we are mostly interested in the decay of a parent-neutrino into a daughter-neutrino (or an antineutrino into an antineutrino) and will concentrate on the Dirac case henceforth. There are some reasons for this. One is that the decay of a parent-neutrino into a daughter-antineutrino leads to qualitatively different phenomenology since it is often the case that the detector response to neutrinos is very different from the response to antineutrinos. The other is that we ultimately want to identify effective-theory descriptions of neutrino decay that are mostly unconstrained by experimental probes other than neutrino decay.  This is not the case of Eq.~(\ref{eq:maj_su2}).

If the neutrinos are Dirac fermions, Eq.~(\ref{eq:maj_su2}) may still be present as long as $\varphi$ carries lepton number $-2$ (for more details, see Ref.~\cite{Berryman:2018ogk}). In this case, however, only neutrino to antineutrino decays are present. Instead, if $\varphi$ caries zero lepton number, Eq.~(\ref{eq:maj_su2}) is forbidden\footnote{The same applies to new interactions proportional to $\nu^c\nu^c\varphi^*$, which are also alllowed if $\varphi$ carries lepton number $-2$.} and two-body neutrino decays are mediated by the dimension-five
\begin{equation}
{\cal L}_{\rm 2b-decay} \propto (HL)\nu^c\varphi + h.c.,
\label{eq:dir_su2}
\end{equation} 
where $\nu^c$ is a two-component left-chiral antineutrino field. After electroweak symmetry breaking, 
\begin{equation}
{\cal L}_{\rm 2b-decay} \propto \nu\nu^c\varphi + h.c.,
\label{eq:dir}
\end{equation} 
which mediates the decays of interest here, $\nu\to\nu+\varphi^*$ and $\bar{\nu}\to\bar{\nu}+\varphi$. Eq.~(\ref{eq:dir_su2}) also mediates processes other than neutrino decay, including $\pi^-\to e^- \bar{\nu}\varphi$ and other interesting meson decays and scattering processes. Nonetheless, the differential decay widths from Eq.~(\ref{eq:dir}), discussed in detail in Sec.~\ref{sec:2body}, will help us understand the impact of nonzero daughter masses in the laboratory frame and will provide some intuition for interpreting the more complicated three-body decays. 

\subsection{Three-Body-Decay Interactions}

Assuming there are no new particles lighter than the known neutrinos, the decay of one heavy neutrino into three lighter ones is mediated by four-fermion operators. There are, if the neutrinos are Dirac fermions, three types of dimension-six operators, defined by how many $L$ and $\nu^c$ fields they contain.\footnote{We do not consider operators with mass-dimension eight or higher.} Schematically, they are proportional to $(LL)^{\dagger}LL$, $(L\nu^c)^{\dagger}L\nu^c$, and $(\nu^c\nu^c)^{\dagger}\nu^c\nu^c$. While the first two mediate strongly constrained processes involving charged leptons ($\mu^+\to e^+e^-e^+$, extra contributions to $\mu^{\pm}\to e^{\pm}\bar{\nu}\nu$, etc) the ``all-singlets'' operator involves only left-chiral antineutrinos. The interactions of left-handed antineutrinos are virtually unconstrained thanks to the fact that neutrino masses are tiny and all neutrinos are ultrarelativistic in the laboratory frames of all experiments executed to date. In summary, we will concentrate on 
\begin{equation}
{\cal L}_{\rm 3b-decay} = -\frac{1}{\Lambda_{\nu}^2}\nu^c\nu^c(\nu^c\nu^c)^{\dagger},
\label{eq:3b}
\end{equation} 
where $\Lambda_{\nu}$ is a constant with dimensions of mass. It is straight-forward to estimate the neutrino lifetime associated to Eq.~(\ref{eq:3b}). Ignoring phase-space effects related to the masses of the daughter-neutrinos,
\begin{equation}
\tau_{\nu} \sim \left(\frac{\Lambda_{\nu}}{\rm eV}\right)^4 \left(\frac{10^{-1}\rm eV}{m_{\nu}}\right)^5\times 10^{-5}~\rm s,
\end{equation}
where $m_{\nu}$ is the mass of the parent-neutrino. Hence, eV-scale new interactions among the left-handed antineutrinos lead to neutrino lifetimes which are of order 10~microseconds. These are within reach of long-baseline, solar-system-bound experiments. For example, for $E_{\nu}\sim 5$~MeV, typical of reactor antineutrinos and high-energy solar neutrinos, the lab-frame decay length of a neutrino with $m_{\nu}=10^{-1}$~eV, for $\Lambda_{\nu}=1$~eV, is $\gamma c\tau\sim 10^8$~km, of order the Earth--Sun distance. 

\section{Neutrino decay: Propagation and Detection}
\label{sec:formalism}
\setcounter{equation}{0}

In this section, we review how neutrino decay impacts neutrino flavor-evolution as a function of baseline. First, we review the formalism for invisible daughters followed by a discussion of decays into visible products.

\subsection{Invisible Decays}

Invisible decays refer to scenarios where the decay products do not interact, for all practical purposes, with the detector. This is the case, for example, when the neutrino decays into exotic new particles with no or very suppressed SM interactions or into SM particles with energies below detection threshold.   

We consider the decay of one of the massive neutrinos $\nu_h$, i.e., $\nu_h \to {\rm invisible}$. In the case of the normal mass ordering for the neutrinos, both $\nu_2$ and $\nu_3$ are guaranteed to be massive, while in the inverted-mass-ordering case both $\nu_1$ and $\nu_2$ are guaranteed to be massive. For invisible decays, the probability of obtaining a neutrino flavor $\nu_\beta$ starting from a neutrino flavor $\nu_\alpha$ a distance $L$ away is ~\cite{Lindner:2001fx}
\begin{eqnarray}
\label{eqn:invis}
    P^{\rm invisible}_{\nu_\alpha\rightarrow\nu_\beta (L)}=\left|\sum_{i=1}^3 U_{\alpha i}\,U_{\beta i}^*\, {\rm exp}\left(-i\frac{m_i^2L}{2E}\right){\rm exp}\left(-\delta_{ih}\frac{m_i\Gamma_h\,L}{2E}\right)\right|^2\,,
\end{eqnarray}
where $m_i$ is the mass of the $i$th neutrino mass eigenstate, $\Gamma_h$ is the total decay width of $\nu_h$ and $E$ is the neutrino energy. $U_{\alpha i}$, $\alpha=e,\mu,\tau$ are the elements of the neutrino mixing matrix. 

A nonzero $\Gamma_h$ leads to an exponential decay of the $\nu_h$ component of the propagating neutrino relative to the $\Gamma_h\to 0$ limit and modifies the ``amount'' of neutrinos reaching the detector. It is trivial to generalize Eq.~(\ref{eqn:invis}) and allow for two or all of the neutrinos to decay invisibly, each with a different decay width $\Gamma_i$.

\subsection{Visible Decay into Active Neutrinos}

Visible decays refer to scenarios where some of the daughter neutrinos interact with the detector as efficiently as their parents. In this case, the effect of the decay can lead to an extra contribution to the number of events in the detector which depends on the flavor composition of the decay products. 

Let us consider the visible decay of the  heavy neutrino state $\nu^r_h$ with mass $m_h$ and helicity $r$ into a lighter neutrino mass-eigenstate $\nu^s_l$ with mass $m_l$ and helicity $s$, i.e.,  $\nu^r_h \to \nu^s_l + \varphi$.\footnote{In the case of the normal mass ordering, the kinematically allowed decays are $3\to 2$, $3\to 1$, and $2\to 1$. In the case of the inverted mass ordering, they are $2\to 1$, $2\to 3$, $1\to 3$.} The daughter neutrino can be produced at any distance $L_l$ from the parent-neutrino source but we only consider $L_l$ values that are smaller than the baseline $L$, taking into account that the parent neutrinos are ultrarelativistic and hence their daughters tend to decay predominantly in the very forward direction.  Momentum conservation dictates that the energy $E_l$ of $\nu_l$ lies in the range 
\begin{equation}
E_l\in\left[\left(\frac{m_l^2}{m_h^2}\right) E_h, E_h\right].
\label{eq:E_lrange} 
\end{equation}
Assuming that $\nu_l$ is itself stable, the differential probability\footnote{This object is not, strictly speaking, a probability since, for example, it is not constrained to be less than one. Keeping this in mind, however, it can be used, when computing the expected number of events as a function of energy in a concrete experimental setup, just like the oscillation probability as long as some care is given to cross-section-related issues, to be discussed momentarily.} of obtaining a neutrino flavor $\nu_\beta$ starting from another flavor $\nu_\alpha$, a distance $L$ away, is~\cite{Lindner:2001fx,Coloma:2017zpg,Gago:2017zzy,Porto-Silva:2020gma},
\begin{eqnarray}
\label{eqn:vis}
   \frac{dP_{{\nu^r_\alpha\rightarrow\nu^s_\beta}(L)}}{dE_l}&=&
   \left|\sum_{i=1}^3 U_{\alpha i}\,U_{\beta i}^*\, {\rm exp}\left(-i\frac{m_i^2L}{2E_h}\right){\rm exp}\left(-\delta_{ih}\frac{m_i\Gamma_h\,L}{2E_h}\right)\right|^2\, \delta(E_h-E_l) \, \delta_{rs}\nonumber\\
   &+& 
   \eta^{rs}_{hl}(E_h,E_l)\vert U_{\alpha h}\vert^2\vert U_{\beta l}\vert^2\left[1-{\rm exp}\left(-m_h\Gamma_h\,\frac{ L}{E_l}\right)\right]\,.
\end{eqnarray}
Here $\eta^{rs}_{hl}(E_h,E_l)$ is the differential decay width of the different helicity contributions $(\Gamma^{rs})$, normalized to the total decay width $\Gamma_{h}(E_h)$, both evaluated in the lab frame: 
\begin{equation}
\label{eq:eta}
     \eta^{rs}_{hl}(E_h,E_l)=\frac{1}{\Gamma_h (E_h)}\left[\frac{d\Gamma^{rs}}{dE_l}(E_h,E_l)\right].
\end{equation}
Depending on the neutrino source, $\eta^{rs}_{hl}(E_h,E_l)$ should also contain a geometrical factor that includes the probability that the daughter-neutrino direction is such that it ends up in the detector; for a detailed discussion, see~\cite{Lindner:2001fx}. We do not worry about this here because we will be interested in a practically isotropic neutrino source -- in Sec.~\ref{sec:JUNO}, a nuclear reactor -- and will also take advantage of the fact that the neutrino decays are very forward, as will be discussed in Sec.~\ref{sec:2body}.
It is evident from Eq.~(\ref{eqn:vis}) that the net probability has two contributions: the first line contains the probability of obtaining $\nu_\beta$ from the surviving parent, whereas the second term is the probability of obtaining, when the decay occurs, $\nu_\beta$ from the daughter produced in the decay. 

Eq.~(\ref{eqn:vis}) does not include cross-section-related information, which may be different for parent and daughters. In particular, we allowed for both left-handed and right-handed helicity daughters and, given the parity-violating nature of the weak interaction, neutrinos with different helicities interact very differently. In fact, neutrino decays can be safely considered invisible when the daughter-neutrinos have the ``wrong helicity,'' i.e., when the daughter-neutrinos are right-handed or the daughter antineutrinos are left-handed. This is an excellent approximation when neutrinos are ultrarelativistic in the lab-frame and assuming neutrino detection is mediated by the weak interactions. Under these circumstances, the neutrino-detection cross sections are suppressed by a chirality violating factor of order $(m_{\nu}/E)^2$. After taking cross-section information into account, for wrong-helicity daughters, Eq.~(\ref{eqn:vis}) is equivalent, when it comes to making any useful estimate, to Eq.~(\ref{eqn:invis}).

The energy distribution of the daughter neutrinos is encoded in $\eta^{rs}_{hl}(E_h,E_l)$, and depends on the helicities of the parent and the daughter. It also depends on the parent and daughter masses and on the nature of the interaction responsible for neutrino decay. We will discuss this in detail in the context of specific scenarios in Sections~\ref{sec:2body} and \ref{sec:3body}.

\section{Two-Body Neutrino Decay}
\label{sec:2body}
\setcounter{footnote}{0}
\setcounter{equation}{0}

In this section, we compute the differential decay width for the two-body decay $\nu_h\rightarrow\nu_l+\varphi$ for different neutrino-decay scenarios. We assume that $\nu_h$ and $\nu_l$, with masses $m_h$ and $m_l<m_h$, are mass eigenstates and linear superpositions of the three active flavor eigenstates, $\nu_e,\nu_{\mu},\nu_{\tau}$. As already mentioned, we will consider the neutrinos to be Dirac fermions so lepton number is conserved and assume that $\varphi$ carries zero lepton number. 

As discussed in Sec.~\ref{sec:models}, we are interested in models where neutrino decay is mediated by Eq.~(\ref{eq:dir}). Here we rewrite Eq.~(\ref{eq:dir}), including coupling constants, labeling the different neutrino states, and expressing those as four-component Dirac fields:
\begin{equation}
\label{eqn:L}
\mathcal{L} \supset g_\varphi\,\overline{\nu}_l\left(\A\,\mathbb{P_R}+\B\,\mathbb{P_L}\right)\nu_h\,\varphi+h.c.\,,
\end{equation}
where $\mathbb{P_{R,L}}=(1\pm \gamma_5)/2$ are the chirality-projection operators, and $g_\varphi,\A,\B$ are dimensionless constants. $\A,\B$ are normalized such that $|\A|^2+|\B|^2=1$. Since we will be discussing reactor antineutrino experiments in Sec.~\ref{sec:JUNO}, we will focus on the decays of antineutrinos but the discussion can be trivially extended to neutrino decays.

In the rest frame of the parent-antineutrino, the decay width is
\begin{equation}
\Gamma_h = \frac{g_{\varphi}^2m_h}{32\pi}\left(1+\frac{m_l^2}{m_h^2}\right)\left(1-\frac{m_l^2}{m_h^2}\right).
\label{eq:Gtot}
\end{equation}
In the laboratory frame, $\Gamma_h$ is ``boosted'' by a factor $m_h/E_h$:   $\Gamma_h(E_h)= \Gamma_h m_h/E_h$.\footnote{We will usually refer to the decay width $\Gamma$ as the decay width computed in the rest frame of the parent, as is customary. We will flag the decay width in the lab frame explicitly (e.g.,~$\Gamma_h(E_h)$). When there is the possibility of confusion, or when we are working explicitly in the rest frame, we will refer to the decay width in the rest-frame as $\Gamma^*$.} 
This is the quantity used to compute the normalized differential partial widths, Eq.~(\ref{eq:eta}), in the laboratory frame.

In the lab frame, the differential partial width for a $\overline{\nu}_h$ with helicity $r$ decaying into a $\overline{\nu}_l$ with helicity $s$ is
\begin{equation}
\label{eqn:dw}
\frac{d\Gamma_{rs}}{dE_l}=\frac{1}{16\pi E_h p_h}|\mathcal{M}_{rs}|^2\,,
\end{equation}
where the magnitude of the amplitude squared is, when either $\A$ or $\B$ vanish
\begin{equation}
\label{eqn:mat}
|\mathcal{M}_{rs}|^2 = \frac{1}{2}|g_\varphi|^2\left(1-(-1)^{\delta_{rs}}\cos\theta_l\right) \, \times
\left\{
\begin{array}{l@{\quad}l}
\left(E_h-(-1)^{\delta_{1r}}\sqrt{E_h^2-m_h^2}\right)\left(E_l-(-1)^{\delta_{-1s}}\sqrt{E_l^2-m_l^2}\right)\,, & \mathcal{A}=0 \\
\left(E_h-(-1)^{\delta_{-1r}}\sqrt{E_h^2-m_h^2}\right)\left(E_l-(-1)^{\delta_{1s}}\sqrt{E_l^2-m_l^2}\right)\,, & \mathcal{B}=0
\end{array} \right. \,,
\end{equation}
for $r{\rm ,}s\in \{1,-1\}\equiv$ \{right-handed helicity,~left-handed helicity\}. $\theta_l$ is the angle defined by the direction of the three-momentum $\mathbf{p}_l$ of $\overline{\nu}_l$ relative to the direction of the three-momentum $\mathbf{p}_h$ of $\overline{\nu}_h$, 
\begin{equation}
\label{cos}
    \cos\theta_l=  \frac{\mathbf{p}_h\cdot\mathbf{p}_l}{|\mathbf{p}_h||\mathbf{p}_l|} = \frac{1}{|\mathbf{p}_h||\mathbf{p}_l|}\left(E_hE_l-\frac{m_h^2 + m_l^2}{2}\right)\,.
\end{equation}
In the case of two-body decays, $\cos\theta_l$ is not an independent variable; it is a function of $E_l$.

Depending on whether the decay channel is helicity-preserving $(r=s)$, or helicity-flipping $(r\neq s)$, the matrix-element squared (Eq.~(\ref{eqn:mat})) is proportional to $(1\pm \cos\theta_l)$. This can be understood from angular momentum conservation. In the laboratory frame, the daughter with the same helicity as the parent is more likely to be emitted in the direction of the parent (hence vanishes at $\theta_l=\pi$), whereas the daughter with the opposite helicity prefers to be emitted in the opposite direction (hence vanishes at $\theta_l=0$). 
In the lab frame, the daughter is emitted in a narrow forward cone, i.e.,  $\cos\theta_l\approx 1$ (see Eq.~(\ref{cos})). As $m_l$ approaches $m_h$, the opening angle of the cone decreases  ($\cos\theta_l\rightarrow 1$) and $E_l\rightarrow E_h$. In this limit, the $r=-s$ partial width vanishes by angular-momentum conservation. This means that, in the limit $m_l\to m_h$, the helicity-flipping decay is suppressed relative to the helicity preserving one.\footnote{In the limit $m_l\to m_h$, of course, the branching ratio itself is phase-space suppressed ($\Gamma_h\to 0$). We are not concerned about this here and are more interested with the relative sizes of the helicity-flipping and helicity-preserving contributions.}

On the other hand, when the daughter is massless, the matrix-elements squared in Eq.~(\ref{eqn:mat}) vanish for some values of the daughter helicity independent from that of the parent. The reason is that, in the massless limit, the chirality eigenstates $\nu_R$ and $\nu_L$ are exclusively responsible for creating and destroying, respectively, left-handed and right-handed antineutrino daughters. Hence, when $\A=0$, only left-handed antineutrino daughters ($s=-1$) can be produced, while only right-handed antineutrino daughters ($s=1$) can be produced when $\B=0$.  
    
The differential decay-width is easy to describe in the parent rest-frame. Defining $\theta_{\star}$ to be the rest-frame angle between the polarization of the parent particle, assumed to be 100\% polarized, and the direction of the daughter momentum,\footnote{The helicity of the parent at rest is ill-defined, and must be replaced by its polarization. An arrow has been used to indicate the polarization of the parent (in this case, in the positive $\hat{z}$-direction).} 
\begin{equation}
\frac{1}{\Gamma_{\uparrow s=\pm1}}\frac{d\Gamma_{\uparrow s=\pm 1}}{d\cos\theta_{\star}}=1\pm\cos\theta_{\star},
\end{equation}
 independent from the mass of the daughter . When the daughter is massless, its energy in the rest-frame is $m_h/2$ and, in the laboratory frame, taking advantage of the fact that $E_h/m_h\gg 1$,
\begin{equation}
\cos\theta_{\star}=2\left(\frac{E_l}{E_h}\right)-1.
\end{equation}
With this information, it is easy to show that, in the limit $m_l\to 0$ and $E_h/m_h\gg 1$, the differential partial widths have a simple linear form, proportional to either $E_l/E_h$ or $1-E_l/E_h$, depending on the relative helicities of the parent-daughter pair. Here, we take advantage of the very important fact that, for massless fermions, the helicity is reference-frame independent and hence a left-handed daughter-neutrino in the rest frame is left-handed in the laboratory frame. 

In summary, when $m_l=0$, $E_h/m_h\gg 1$ and $\A =0$,
\begin{eqnarray}
\frac{E_h}{\Gamma}\frac{d\Gamma_{1,1}}{dE_l} &=&0, \nonumber \\
\frac{E_h}{\Gamma}\frac{d\Gamma_{1,-1}}{dE_l} &=& 2\left(1-\frac{E_l}{E_h}\right), \nonumber \\
\frac{E_h}{\Gamma}\frac{d\Gamma_{-1,1}}{dE_l} &=& 0, \label{eq:A=0} \\
\frac{E_h}{\Gamma}\frac{d\Gamma_{-1,-1}}{dE_l} &=& 2\frac{E_l}{E_h}.  \nonumber
\end{eqnarray}
Here, $\Gamma$ is the total decay width which, of course, does not depend on $r$ -- a particle's lifetime does not depend on its polarization state! -- or $s$, which is summed over. Eqs.~(\ref{eq:A=0}) state that, for massless daughters, antineutrinos decay only into left-handed antineutrinos and hence, in this limit, the decay-daughters are always invisible, independent from the helicity of the parent.  

Similarly, when $m_l=0$, $E_h/m_h\gg 1$ and $\B=0$,
\begin{eqnarray}
\frac{E_h}{\Gamma}\frac{d\Gamma_{1,1}}{dE_l} &=& 2\frac{E_l}{E_h}, \nonumber \\
\frac{E_h}{\Gamma}\frac{d\Gamma_{1,-1}}{dE_l} &=& 0, \nonumber \\
\frac{E_h}{\Gamma}\frac{d\Gamma_{-1,1}}{dE_l} &=& 2\left(1-\frac{E_l}{E_h}\right),  \label{eq:B=0} \\
\frac{E_h}{\Gamma}\frac{d\Gamma_{-1,-1}}{dE_l} &=& 0.  \nonumber
\end{eqnarray}
Unlike the $\A=0$ case, Eqs.~(\ref{eq:B=0}) state that, for massless daughters, antineutrinos decay only into right-handed antineutrinos and hence, in this limit, the decay-daughters are always potentially visible, independent from the helicity of the parent.

\subsection{Differential decay width: massive-daughter}

\begin{figure}[htbp]
\includegraphics[width=1\textwidth]{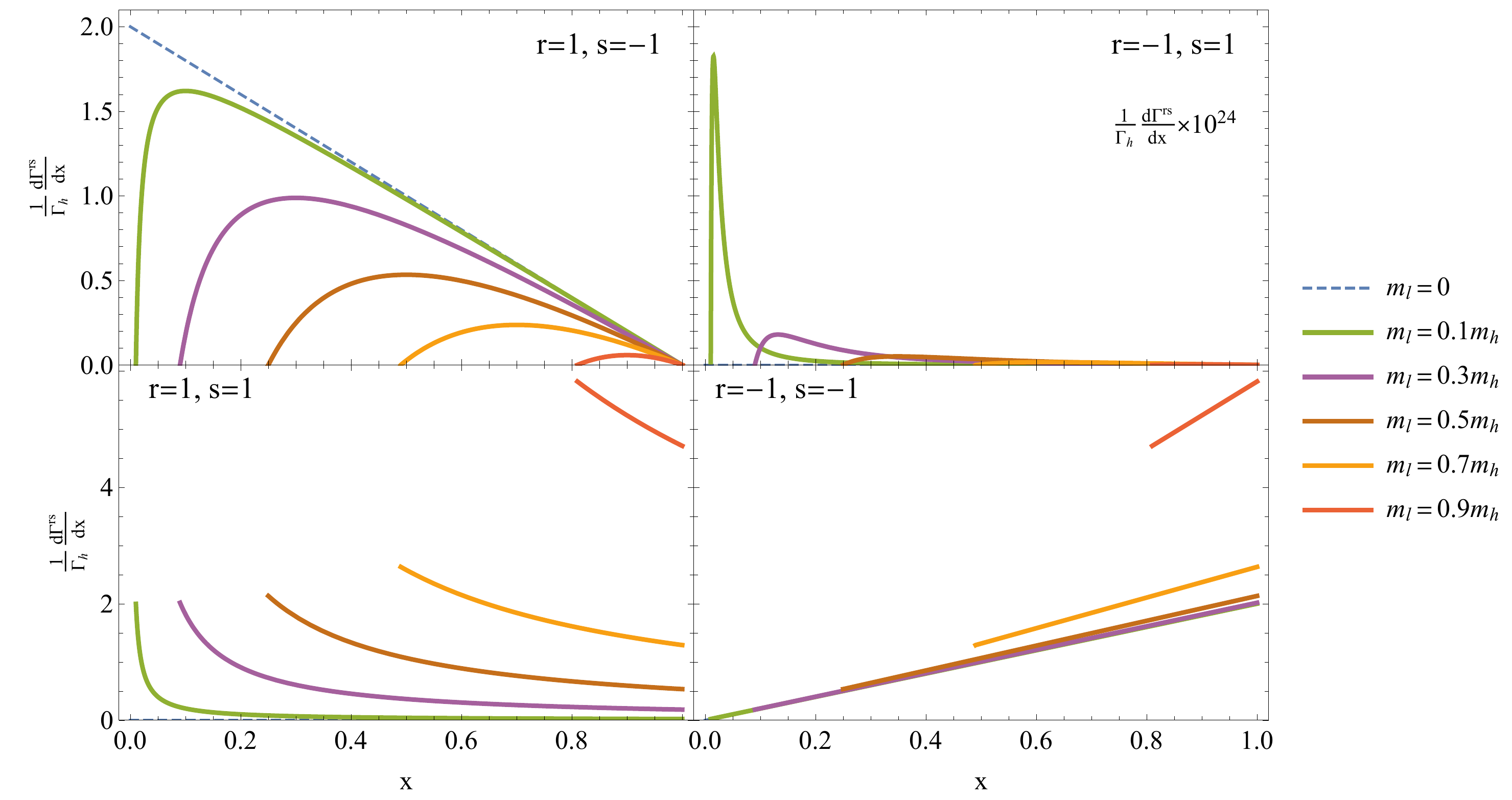}
\caption{Differential decay widths for $\overline{\nu}_h^r\rightarrow\overline{\nu}_l^s+\varphi$ normalized to the total width, as a function of $x=E_l/E_h$ for $\A=0$ (see Eq.~(\ref{eqn:L})) and all possible $r,s=\pm1$ combinations, for different values of the daughter-to-parent mass ratio.}
\label{fig:A0}
\end{figure}

We are especially interested in exploring the consequences of $m_l\neq0$ and how the information in Eqs.~(\ref{eq:A=0}, \ref{eq:B=0}) above, changes. We will often draw attention to the case $\A=0$ when, in the massless-daughter limit, only left-handed antineutrino daughters $(s=-1)$ are produced in the decay of antineutrino parents.  
 Fig.\,\ref{fig:A0} depicts the normalized differential decay widths for all helicity combinations $(r,s=\pm 1)$, for a $\A=0$ and different masses of the daughter antineutrinos, assuming $m_h/E_h\ll 1$. Clearly, the distributions depend strongly on the mass of the daughter antineutrino, especially as it approaches that of the parent.

We comment on the behavior of the different helicity combinations.
    \begin{itemize}
        \item $r=1,s=1$, Fig.\,\ref{fig:A0}(bottom, left):  
        This contribution can be non-zero only if the daughter mass $m_l$ is not zero. On the other hand, as $m_l\to m_h$, the $s=1$ final state is the only allowed daughter-helicity for the $r=1$ decay so this contribution, while naively chirality-disfavored, eventually dominates over the $r=1$, $s=-1$ chirality favored decay mode.
    
         \item $r=1,s=-1$, Fig.\,\ref{fig:A0}(top, left):  Here the parent and daughter helicities are opposite but the $s=-1$ final state is the only one accessible in the massless limit (chirality-favored). For small $m_l$, the behavior matches expectations from the massless limit, Eqs.~(\ref{eq:A=0}) (decreasing straight line). As $m_l$ approaches $m_h$, this decay mode is disfavored relative to the $r=1$, $s=1$ angular-momentum-favored decay mode.

         \item $r=-1,s=-1$, Fig.\,\ref{fig:A0}(bottom, right):  This final-state is both chirality-favored (when $m_l\ll m_h$) and angular-momentum-favored (when $m_l\to m_h$). Mass effects are consistent with expectations from kinematics (e.g., the restriction of the energy range to harder values of $E_l$).        
        
         \item $r=-1,s=1$, Fig.\,\ref{fig:A0}(top, right): 
        This final-state is both chirality-suppressed (when $m_l\ll m_h$) and angular-momentum-disfavored (when $m_l\to m_h$) and always completely subdominant to the $r=-1$, $s=-1$ decay mode.
        
            \end{itemize}

We summarize the results as follows. If the production of the daughter antineutrino is associated to the right-chiral field $\nu_R=\mathbb{P_{R}}\nu$, in the limit where $m_l\ll m_h$, the decay of antineutrinos is, for all practical purposes, into left-helicity antineutrinos and hence invisible. The situation can be significantly different if the daughter-antineutrino mass and that of the parent are similar, $m_l\sim m_h$. In this case, even for right-chiral produced antineutrinos, one ends up with a healthy fraction of potentially visible (right-handed helicity) daughter antineutrinos. Hence, daughter-mass effects are potentially large even in the limit $m_h/E_h, m_l/E_l\ll 1$. When it comes to the shapes and relative sizes of the partial width distributions, the dimensionless parameter that captures the physics, even in the lab frame, is $m_l/m_h$.

These conclusions are relevant in the case that is phenomenologically interesting. Since neutrinos are, for all practical purposes, always produced via weak interactions, in the laboratory frame all antineutrinos are right-handed ($r=1$) while all neutrinos are left-handed ($r=-1$), up to negligible corrections of order $(m_h/E_h)^2$. As  Fig.\,\ref{fig:A0} reveals, if $m_l$ is large enough, the decay into right-handed antineutrinos can dominate over that into left-handed antineutrinos even when $\A=0$. This behavior is easy to see analytically as well. When the parent antineutrino is right-handed ($r=1$), in the ultrarelativistic limit,
\begin{equation}
\label{eqn:approx}
\frac{d\Gamma_{1,s}}{dE_l} = \frac{|g_\varphi|^2m_h^2}{16\pi E_hE_l} \, \times
\left\{
\begin{array}{l@{\quad}l}
m_l^2/m_h^2\,, & s=1 \\
(E_l/E_h-m_l^2/m_h^2)(1-E_l/E_h)\,. & s=-1
\end{array} \right. 
\end{equation}
The chirality-disfavored $s=1$ daughter-helicity is suppressed by $(m_l/m_h)^2$ relative to the chirality-favored $s=-1$ daughter-helicity when $m_l$ is small. As $m_l$ approaches $m_h$, however, $E_l$ approaches $E_h$ (see Eq.~(\ref{eq:E_lrange})) and the chirality-favored $s=-1$ contribution is suppressed relative to the $s=1$ one. 
In Sec~\ref{sec:JUNO}, we explore the consequences of this result by investigating the potential of JUNO to observe visible and invisible antineutrino decays.\\

\begin{figure}[htbp]
\includegraphics[width=1\textwidth]{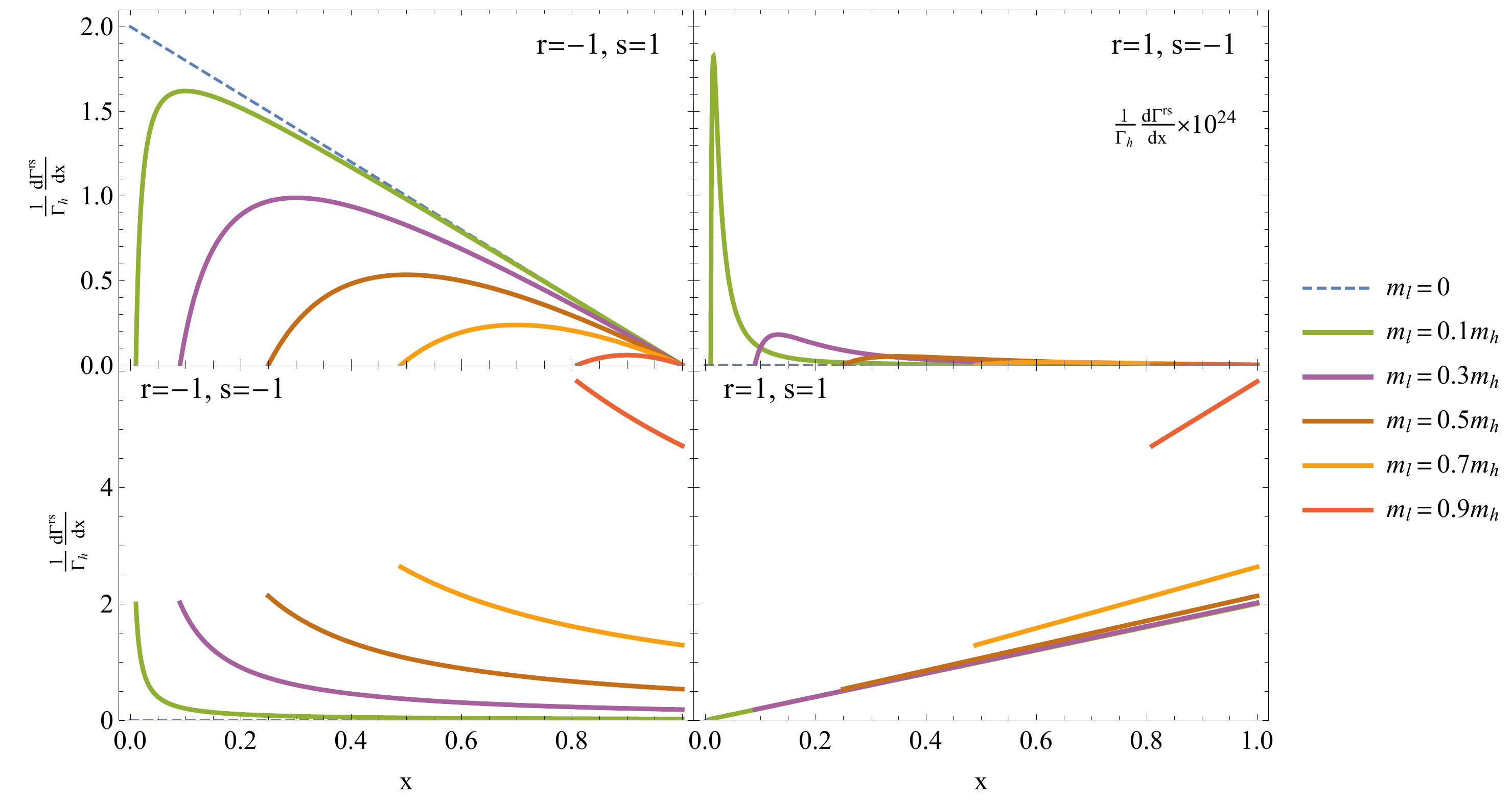}
\caption{Differential decay widths for $\overline{\nu}_h^r\rightarrow\overline{\nu}_l^s+\varphi$ normalized to the total width, as a function of $x=E_l/E_h$ for $\B=0$ (see Eq.~(\ref{eqn:L})) and all possible $r,s=\pm1$ combinations, for different values of the daughter-to-parent mass ratio.}
\label{fig:B0}
\end{figure}

Fig.\,\ref{fig:B0} depicts the shapes of the differential decay widths in the  $\B=0$ case. The results here are the same as those in the $\A=0$ case after one establishes a simple ``map''  between the different parent-daughter helicity combinations: $\{r,s\}\to\{-r,-s\}$. They can be understood by using the same arguments as above. When it comes to phenomenologically interesting scenarios, the $\A=0$ scenario can be very different from the $\B=0$ scenario. When $\B=0$, the daughter-antineutrinos of right-handed parent-antineutrinos are always right-handed, and hence visible, independent from the value of $m_l$.\footnote{There is a $r=1$, $s=-1$ component (left-handed daughters), but it is extremely suppressed (by $10^{-24}$) as is clear from Fig.\,\ref{fig:B0}.}

\section{Neutrino decay at JUNO}
\label{sec:JUNO}
\setcounter{equation}{0}
\setcounter{footnote}{0}

The Jiangmen Underground Neutrino Observatory (JUNO) is a 20kt liquid scintillator neutrino experiment based in China. It is currently under construction and expected to start operations later in 2022. JUNO will be able to measure the neutrino oscillations parameters $\sin^2\theta_{12}$, $\Delta m^2_{21}$ and $|\Delta m^2_{32}|$ with sub-percent precision and will shed new light on the neutrino mass ordering at a $3-4 \sigma$ level after approximately 6 years of data taking~\cite{JUNO:2021vlw}. JUNO detects $\overline{\nu}_e$ primarily through inverse beta decay (IBD), $\overline{\nu}_e+p\to n+e^+$. It determines the neutrino energy by measuring the energy of the recoil positron and can reconstruct the neutrino energy spectrum with very good precision. The $\overline{\nu}_e$ are by-products of the burning of the reactor fuel, composed of $^{235}{\rm U}$, $^{238}{\rm U}$, $^{239}{\rm Pu}$, and $^{241}{\rm Pu}$ in the nearby (${\rm L}\sim 53 {\rm km}$) Yangjian and the Taishan nuclear reactor plants. The $\overline{\nu}_e$ flux from different isotopes can be fit using exponentials of fifth order polynomials in the energy range  $\sim [1.8,8]$~MeV~\cite{Mueller:2011nm}, where $1.8$~MeV is the threshold of the IBD process. While computing the number of antineutrino events, we make use of the following average relative isotope composition for the nuclear fuel: $^{235}{\rm U}$~:~$^{238}{\rm U}$~:~$^{239}{\rm Pu}$~:~$^{241}{\rm Pu}$ = $0.577:0.076:0.295:0.052$.
 
The number of events detected at JUNO will depend on the antineutrino decay-process considered. The differential number of events at JUNO including the presence of potentially visible decay-daughters is, 
\begin{eqnarray}
\label{eqn:Ntot}
    \frac{dN^{rs}_{\overline{\nu}_e\rightarrow\overline{\nu}_e(L)}}{dE_l}=\sigma(E_l)\varphi(E_l)P^{\rm invisible}_{\overline{\nu}_e\rightarrow\overline{\nu}_e(L)}+
    \sigma(E_l)\int^{E_{\rm max}}_{E_l} dE_h\varphi(E_h)\frac{dP^{\rm visible}_{\overline{\nu}_e\rightarrow\overline{\nu}_e(L)}}{dE_l}(E_h,E_l)\,,
\end{eqnarray}
where $E_{\rm max}=E_lm^2_h/m^2_l$ is the kinematic limit, see Eq.~(\ref{eq:E_lrange}), and $\sigma$ is the cross section for IBD. $P^{\rm invisible}_{\overline{\nu}_e\rightarrow\overline{\nu}_e(L)}$ is given by Eq.~(\ref{eqn:invis}) for $\alpha=\beta=e$ while 
\begin{equation}
\frac{dP^{\rm visible}_{\overline{\nu}_e\rightarrow\overline{\nu}_e(L)}}{dE_l} =  \eta^{11}_{hl}(E_h,E_l)\vert U_{e h}\vert^2\vert U_{e l}\vert^2\int_0^L\,dL_l\left[1-{\rm exp}\left(-m_h\Gamma_h\,\frac{ L_l}{E_l}\right)\right]\,,
\end{equation}
is the second line in Eq.~(\ref{eqn:vis}) for $r=s=1$, $\alpha=\beta=e$. 

Here we assume all antineutrino parents are right-handed keeping in mind that the ``wrong helicity'' component is suppressed by a factor $m_h^2/E_h^2$. Similarly, for left-handed antineutrino-daughters, the detection cross section is suppressed by a factor of $m_l^2/E_l^2$, and can be safely set to zero. For the oscillation parameters, we make use of the PDG parameterization for the elements of the mixing matrix and use the following values for the oscillation parameters of interest \cite{ParticleDataGroup:2020ssz}
\begin{equation}
    \begin{array}{c}
       \sin^2(\theta_{12})=0.307; \,\,  \sin^2(\theta_{13})=0.0218; \,\, 
       \Delta m_{21}^2=7.54\times 10^{-5}{\rm eV}^2; \,\, \Delta m_{31}^2=2.47\times 10^{-3}{\rm eV}^2\,.
   \end{array}
   \label{eq:osc_param}
\end{equation}
The quantities of interest here are $m_h\Gamma_h$ and the ratio of the parent mass to the daughter mass. It turns out that these are nicely connected once the mass-squared differences are know. In more detail, one can express the parent and daughter masses as a function of the relevant mass-squared difference and the mass ratio:
\begin{equation}
m_h^2=\frac{\Delta m^2_{hl}}{\left(1-\frac{m_l^2}{m_h^2}\right)},~~~~~
m_l^2=\frac{\Delta m^2_{hl}\left(\frac{m_l^2}{m_h^2}\right)}{\left(1-\frac{m_l^2}{m_h^2}\right)}.
\end{equation}
Furthermore, for fixed $\Delta m^2_{hl}$,
\begin{equation}
m_h\Gamma_h = g_{\varphi}^2\frac{\Delta m^2_{hl}}{32\pi}\left(1+\frac{m_l^2}{m_h^2}\right).
\label{eq:mGD}
\end{equation}
Since $m_l^2/m_h^2\in[0,1]$, the dependency of $m_h\Gamma_h$ on the mass ratio, for fixed mass-squared difference, is rather mild. In fact, $m_h\Gamma_h$ is finite in the limit $m_l\to m_h$ (when the decay phase space vanishes). The reason is that as $\Gamma_h$ goes to zero, $m_h,m_l\to\infty$ in the limit $m_l\to m_h$ and fixed $\Delta m^2_{hl}$, such that the product is finite. Eq.~(\ref{eq:mGD}) also reveals that, for perturbative values of the coupling $g_{\varphi}^2< 4\pi$, $m_h\Gamma_h$ is at most of order $\Delta m^2_{hl}$.

In the following subsections, we discuss the effects of visible and invisible neutrino decays on the expected event rates at JUNO.

\subsection{Invisible Decay}

\begin{figure}[htbp]
\includegraphics[width=1\textwidth]{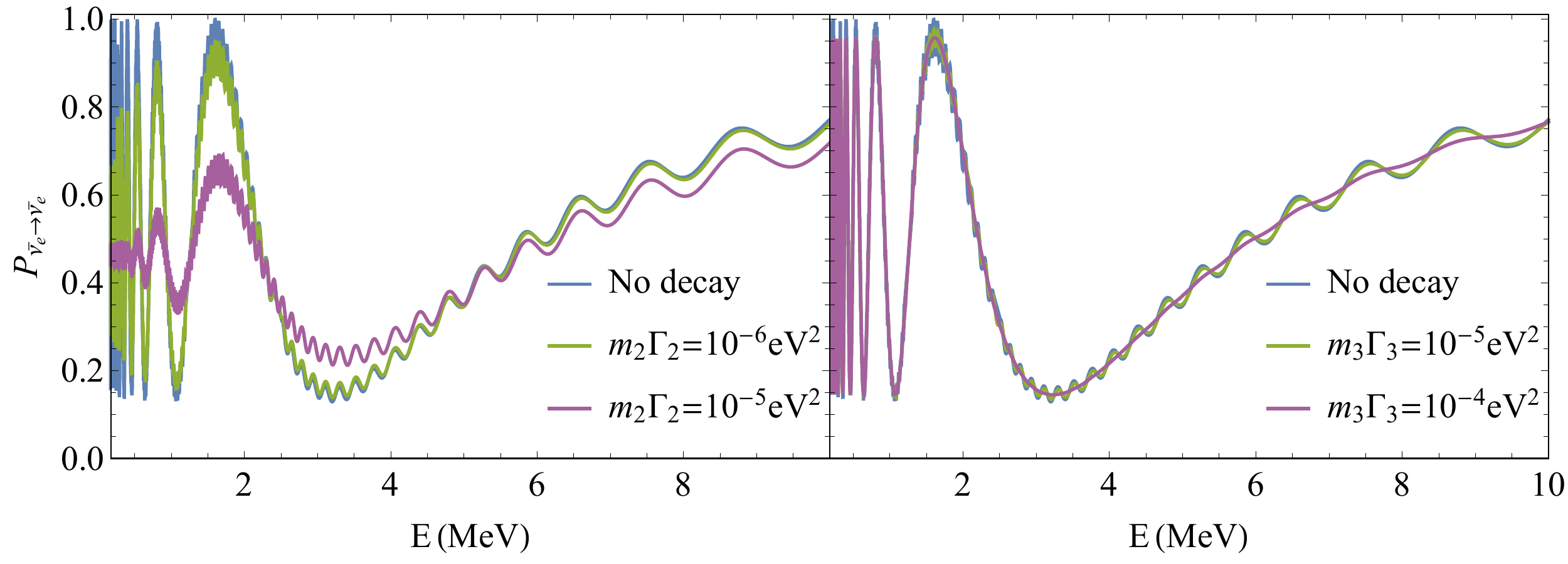}
\caption{$\overline{\nu}_e$ survival probability as a function of the neutrino energy for the JUNO baseline ($L=53$~km) and for different values of $m_h\Gamma_h$, assuming the invisible decay of $\overline{\nu}_2$ (left, $h=2$) and $\overline{\nu}_3$ (right, $h=3$).}
\label{fig:Pinv}
\end{figure}

The effect of the invisible decay of the reactor antineutrinos is captured by Eq.~(\ref{eqn:invis}). Fig.\,\ref{fig:Pinv} depicts $P^{\rm invisible}_{\overline{\nu}_e\rightarrow\overline{\nu}_e}$ for $L=53$~km and different values of $\Gamma_hm_h$ for $\bar{\nu}_2$ decays  (left-hand panel) and for $\bar{\nu}_3$ (right-hand panel). $\Gamma_h$  here refers to the total decay width of the parent antineutrino in its rest frame. 
Invisible neutrino decay leads to the exponential ``disappearance'' of the heavy component of the propagating neutrino. The figure of merit, as can be easily seen in Eq.~(\ref{eqn:invis}), is $m_h\Gamma_h$ and decay effects become visible when $m_h\Gamma_h L/E$ is order one or higher.  $m_h\Gamma_h$ has the same mass-dimension as the neutrino mass-squared differences and plays a comparable role ($m_h\Gamma_h L/E$ versus $\Delta m^2L/E$). Since JUNO is designed to ``see'' mass-squared differences of order $10^{-5}$~eV$^2$ and larger, we anticipate that JUNO will be sensitive to $m_h\Gamma_h\gtrsim 10^{-5}$~eV$^2$. This is indeed what Fig.\,\ref{fig:Pinv} reveals. We note that a similar argument can be made for solar neutrino decay, but with much stronger bounds on the lifetimes of the neutrino mass eigenstates, $m_i\Gamma_i\lesssim 10^{-11}$~eV$^2$ for $i=1,2$ as discussed in \cite{Berryman:2014qha} and $i=3$ as discussed in \cite{Picoreti:2021yct}. This difference in neutrino decay sensitivity between reactor and solar neutrinos is mostly due to the difference in $L/E$, where $L/E\sim 10^4$ $\rm{km}/\rm{GeV}$ for reactor neutrinos and $L/E\sim 10^{11}$  $\rm{km}/\rm{GeV} $ for solar neutrinos. The relative impact of invisible and visible solar neutrino decays on the measurements of $^8\rm{B}$ and $^7\rm{Be}$ neutrinos is the subject of ongoing work by the authors.

More quantitatively, the absence of one of the mass eigenstates leads to the dampening of the effects of one of the two oscillation lengths as the propagating neutrino, which starts out as a linear combination of $\nu_1,\nu_2,\nu_3$, decays into a linear combination of, for example, only the two lightest states. Curiously, there are circumstances when the effect of the neutrino decay is to increase the antineutrino survival probability relative to the case when the neutrino lifetimes are infinite, as one can easily spot in Fig.\,\ref{fig:Pinv}. This is easy to understand. In the absence of neutrino decays, for all $L$ and $E$, $1\ge P_{ee}\ge (|U_{e1}|^2-|U_{e2}|^2-|U_{e3}|^2)^2$, making use of the fact that $|U_{e1}|^2>|U_{e2}|^2+|U_{e3}|^2$. If the $\nu_2$ or $\nu_3$ components decay away, the upper bound decreases and the lower bound increases: 
\begin{equation}
\left(|U_{e1}|^2+|U_{e2}|^2e^{-\frac{m_2\Gamma_2L}{2E}} +|U_{e3}|^2e^{-\frac{m_3\Gamma_3L}{2E}}\right)^2\ge P_{ee}\ge \left(|U_{e1}|^2-|U_{e2}|^2e^{-\frac{m_2\Gamma_2L}{2E}} -|U_{e3}|^2e^{-\frac{m_3\Gamma_3L}{2E}}\right)^2,
\end{equation}
so the envelop of the survival probability is not only shifted down but also ``squeezed'' (exponentially, from high energies to low energies). Finally, invisible decays of $\nu_2$ are proportional to $|U_{e2}|^2\sim 0.3$, while those of $\nu_3$ are proportional to $|U_{e3}|^2\sim 0.02$. Hence, $\nu_2$ decays are more impactful than those of $\nu_3$.

\begin{figure}[!t]
\includegraphics[width=0.48\textwidth]{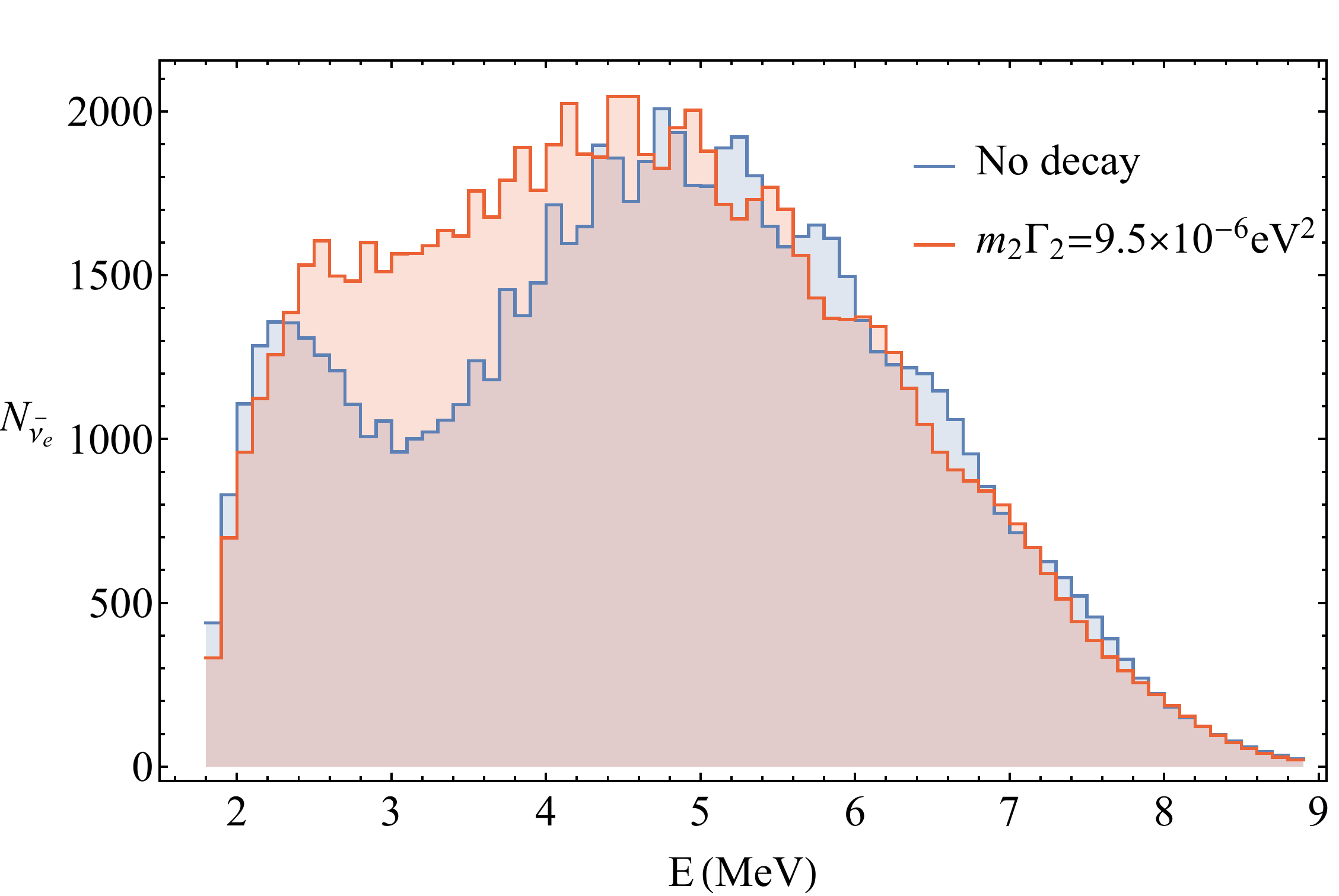}\,\quad\includegraphics[width=0.48\textwidth]{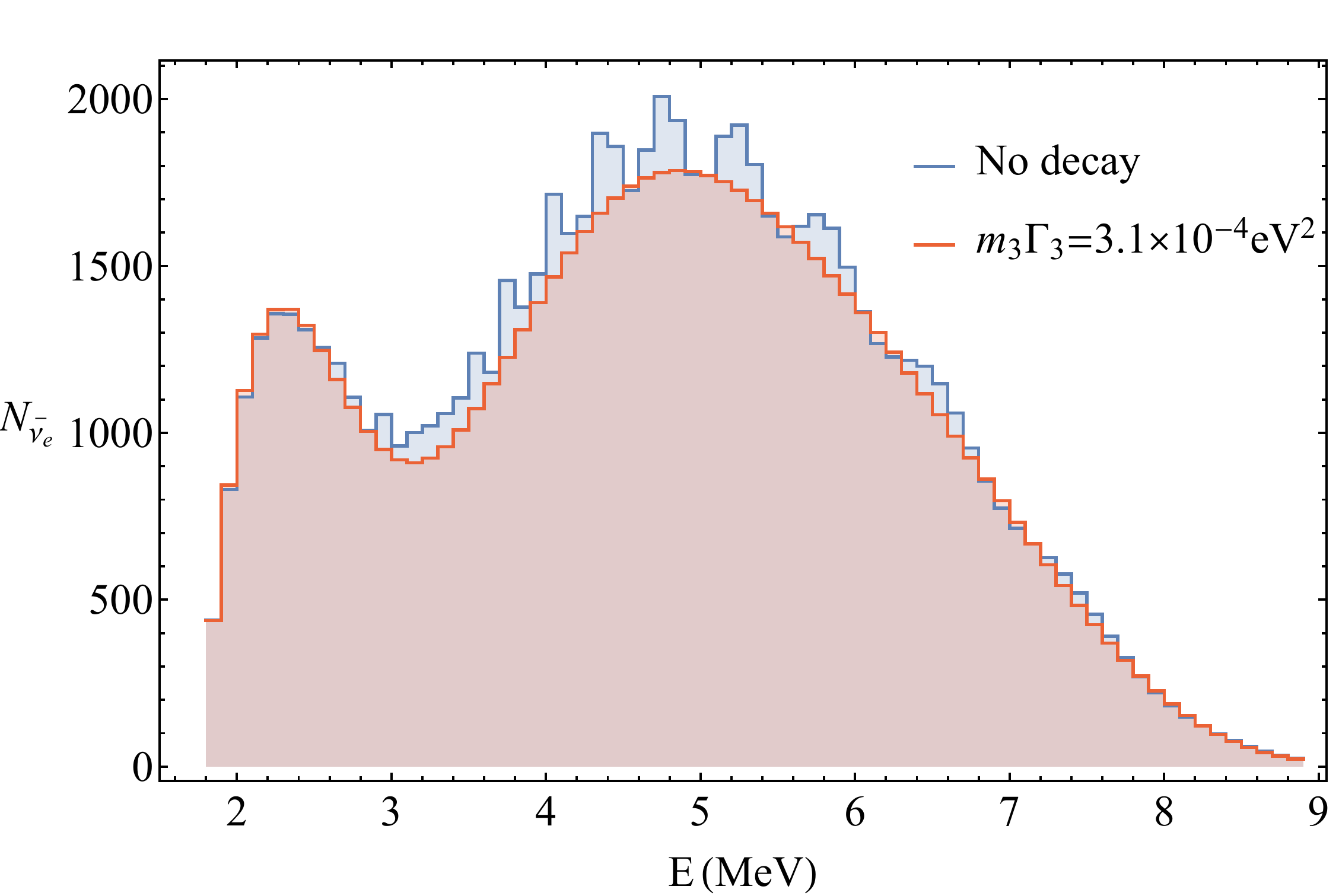}
\caption{Expected number of $\overline{\nu}_e$-mediated events after two years of JUNO data-taking (in energy bins of 0.1~MeV), for different values of $m_h\Gamma_h$, assuming the invisible decay of $\overline{\nu}_2$ (left, $h=2$) and $\overline{\nu}_3$ (right, $h=3$).}
\label{fig:Ninv}
\end{figure}

Fig.\,\ref{fig:Ninv} depicts the expected number of events at JUNO after 2 years of data taking. Events are binned into energy bins of width $0.1$~MeV. The invisible decay of $\bar{\nu}_2$ can lead to a significantly larger number of events relative to the no-decay scenario. In the case of  $\bar{\nu}_3$ decays, we see a slight drop in the total number of events. The specific values of the decay rates are chosen for illustrative purposes; in Sec.~\ref{sec:JUNO_res}, we will present a scan over the decay-parameter space to study the sensitivity to different decay rates.

\subsection{Visible decay}

As mentioned in Sec.~\ref{sec:formalism}, all parent antineutrinos are, in the laboratory frame, right-handed ($r=1$). Furthermore, as discussed in detail in Sec.~\ref{sec:formalism}, if the physics responsible for the decay is captured by Eq.~(\ref{eqn:L}) when $\B=0$, then, independent from the daughter-antineutrino mass, all decay-daughters are also right-handed ($r=1$) and hence `visible.' Instead, if the physics responsible for the decay is captured by Eq.~(\ref{eqn:L}) when $\A=0$, decay-daughters are a combination of visible (right-handed ($s=1$)) and invisible (left-handed (s=-1)). The visible to invisible ratio depends strongly on the mass of the daughter antineutrino. Here we consider the consequences of both interaction scenarios ($\A=0$ and $\B=0$) for JUNO, for different values of $m_l$. For both cases, we consider two independent decay channels, $\overline{\nu}_2\rightarrow\overline{\nu}_1+\varphi$ and $\overline{\nu}_3\rightarrow\overline{\nu}_1+\varphi$. The latter only occurs for the normal mass ordering.\footnote{We don't explore $2\to3$ and $1\to 3$ (inverted mass ordering) or $3\to 2$ decays (normal mass ordering). For these decay modes, the impact of the decay, and the visible daughters, is less pronounced since $|U_{e1}|^2>|U_{e2}|^2>|U_{e3}|^2$.}

The impact of visible decays is captured by Eq.~(\ref{eqn:vis}). In particular, the second line contains the potential contribution from the visible daughters. In this case, we always expect, for the same value of $m_h\Gamma_h$ and the oscillation parameters, more events in the visible case relative to the invisible one. We also expect the ``extra'' events to be softer since the daughter energy is always less than that of the parent. This effect depends on the mass of the daughter antineutrino. 
 
\begin{figure}[t]
\includegraphics[width=0.48\textwidth]{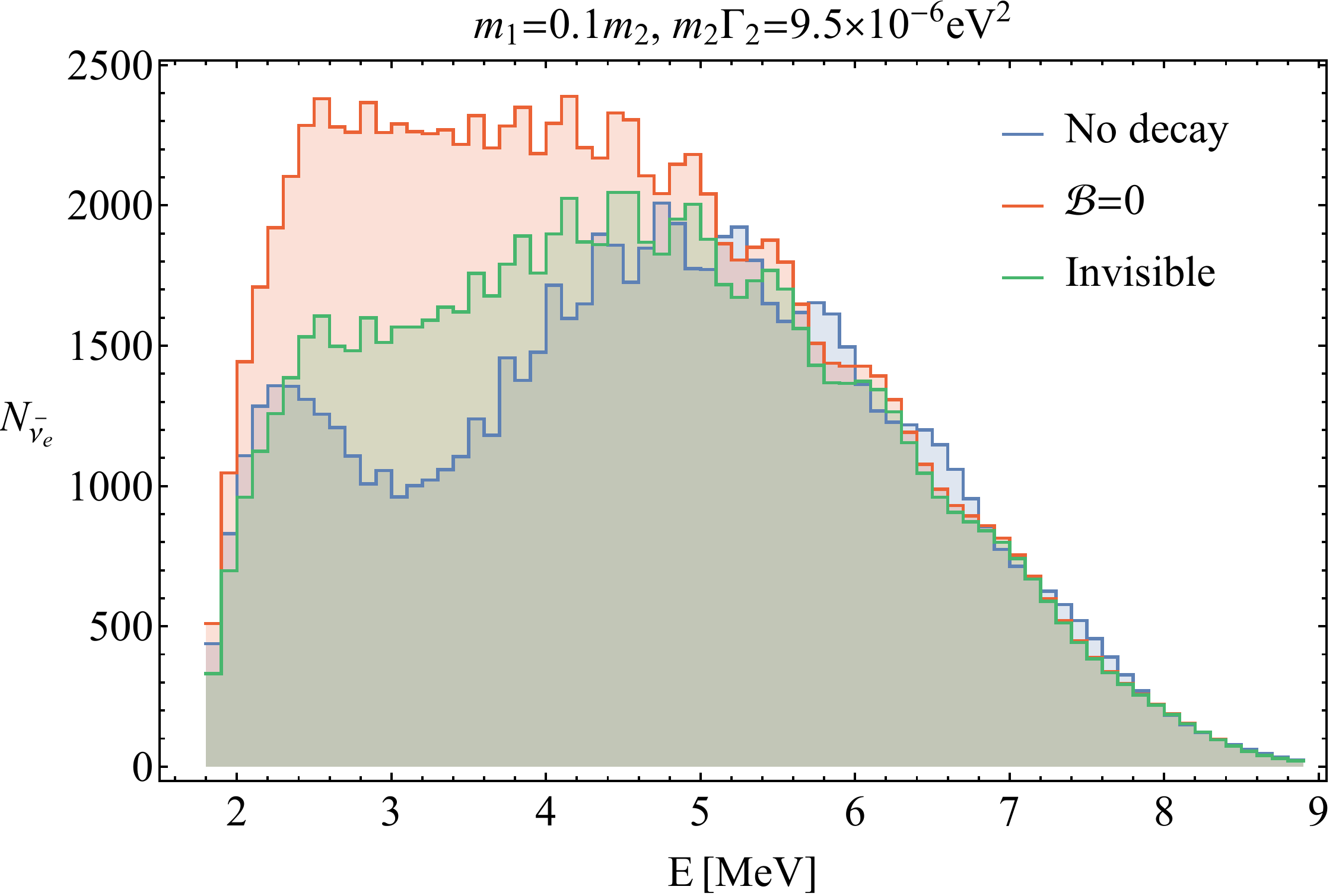}\,\quad
\includegraphics[width=0.48\textwidth]{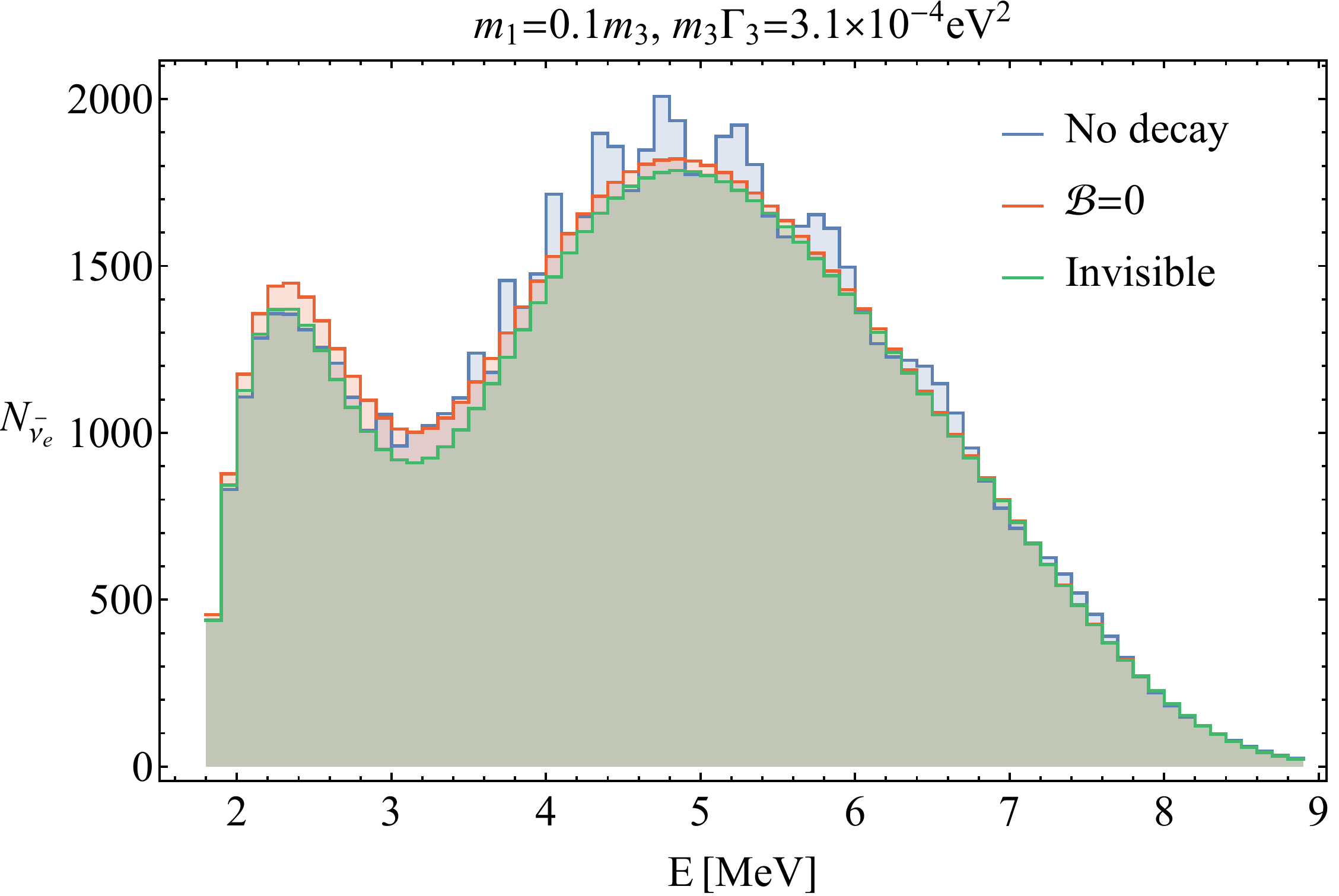}\,
\includegraphics[width=0.48\textwidth]{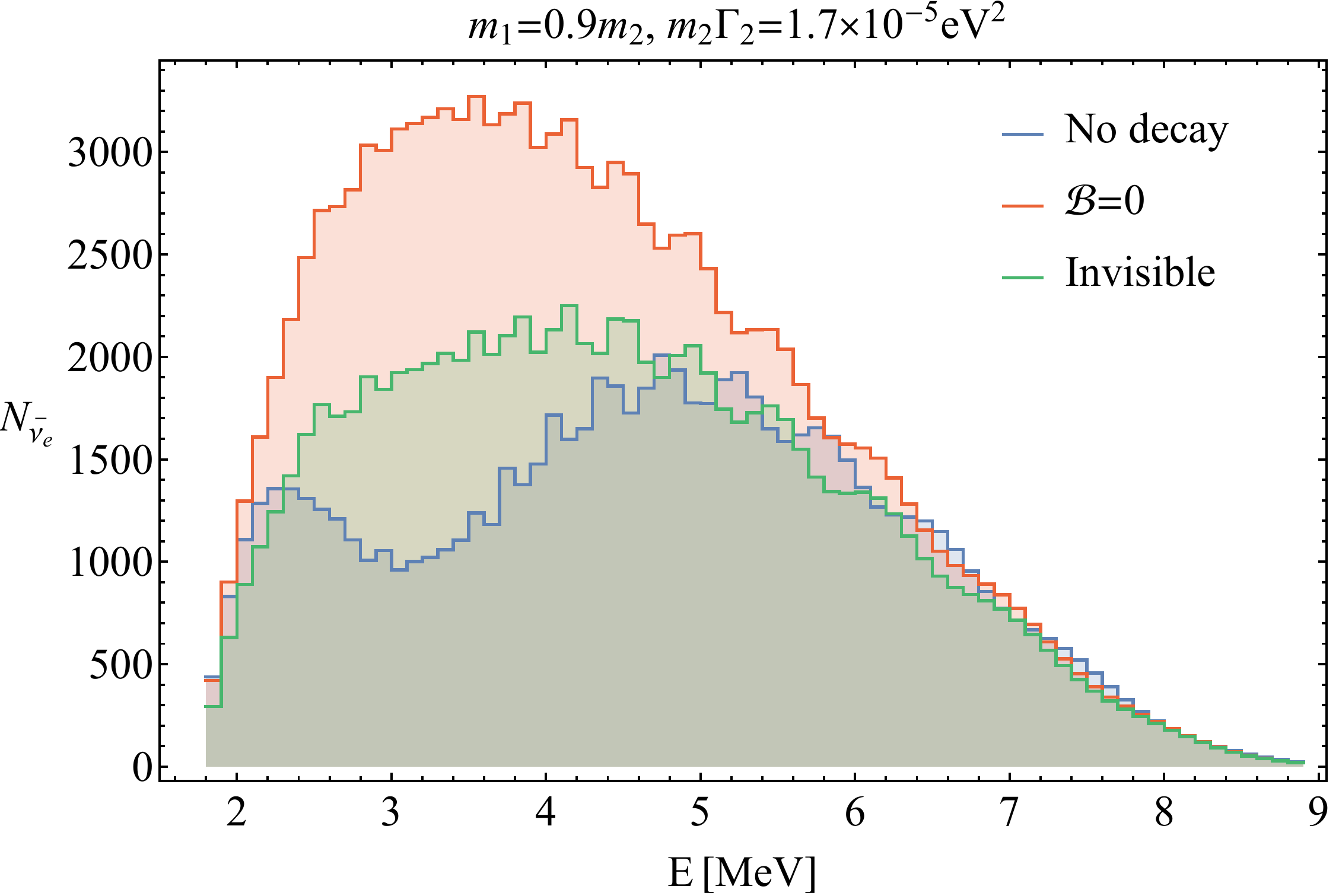}\,\quad
\includegraphics[width=0.48\textwidth]{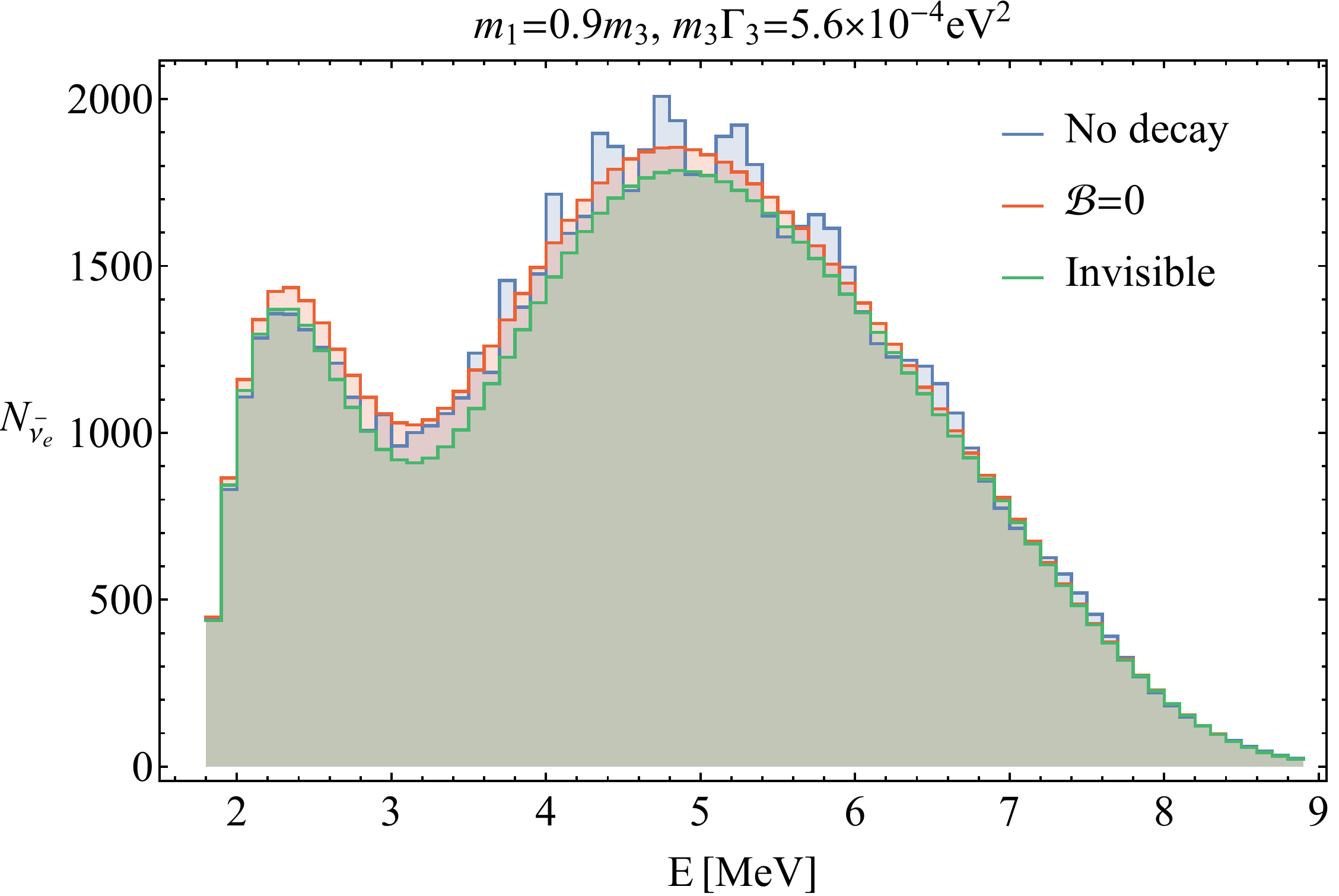}\,
\caption{Expected number of $\overline{\nu}_e$-mediated events after two years of JUNO data-taking (in energy bins of width 0.1~MeV), for different values of $m_h\Gamma_h$ and the ratio of the daughter-to-parent neutrino masses, assuming the neutrino decay is mediated by Eq.~(\ref{eqn:L}) with $\B=0$, for the decays $\overline{\nu}_2\rightarrow\overline{\nu}_1+\varphi$ (left) and $\overline{\nu}_3\rightarrow\overline{\nu}_1+\varphi$ (right) and the corresponding invisible decays.}
\label{fig:NVisB0}
\end{figure}

When $\B=0$, all decay-daughters are visible, independent from the daughter-antineutrino mass (see Fig.\,\ref{fig:B0}). Fig.\,\ref{fig:NVisB0} depicts the expected numbers of events at JUNO after 2 years of data taking where events are binned into energy bins of width $0.1$~MeV for $\overline{\nu}_2\rightarrow\overline{\nu}_1+\varphi$ (left-hand) and  $\overline{\nu}_3\rightarrow\overline{\nu}_1+\varphi$ (right-hand) panels, for different values of the daughter-antineutrino mass. The number of extra events is larger for the $\overline{\nu}_2$ decays relative to those from $\overline{\nu}_3$ decays. This is because the daughter contribution is proportional to $|U_{eh}|^2|U_{el}|^2$, and is therefore much smaller for $\overline{\nu}_3$ decays. Decay parameters are chosen such that, for the two different values of the daughter-to-parent mass ratio, the $g_{\varphi}^2$ is the same. The figure also reveals that, when $m_l\ll m_h$, the daughter energy spectrum is softer. 

\begin{figure}[htbp]
\includegraphics[width=0.48\textwidth]{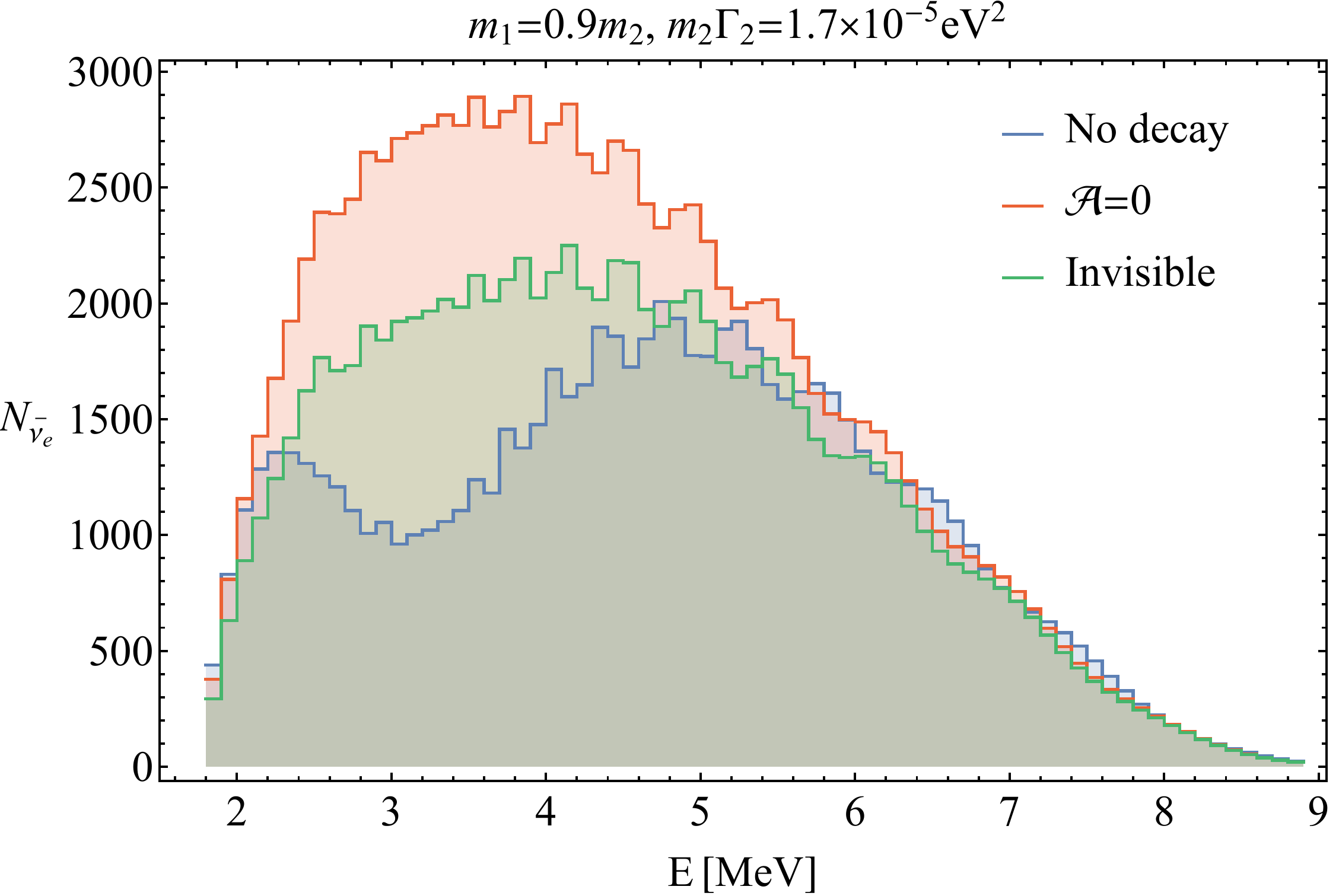} \,\quad\includegraphics[width=0.48\textwidth]{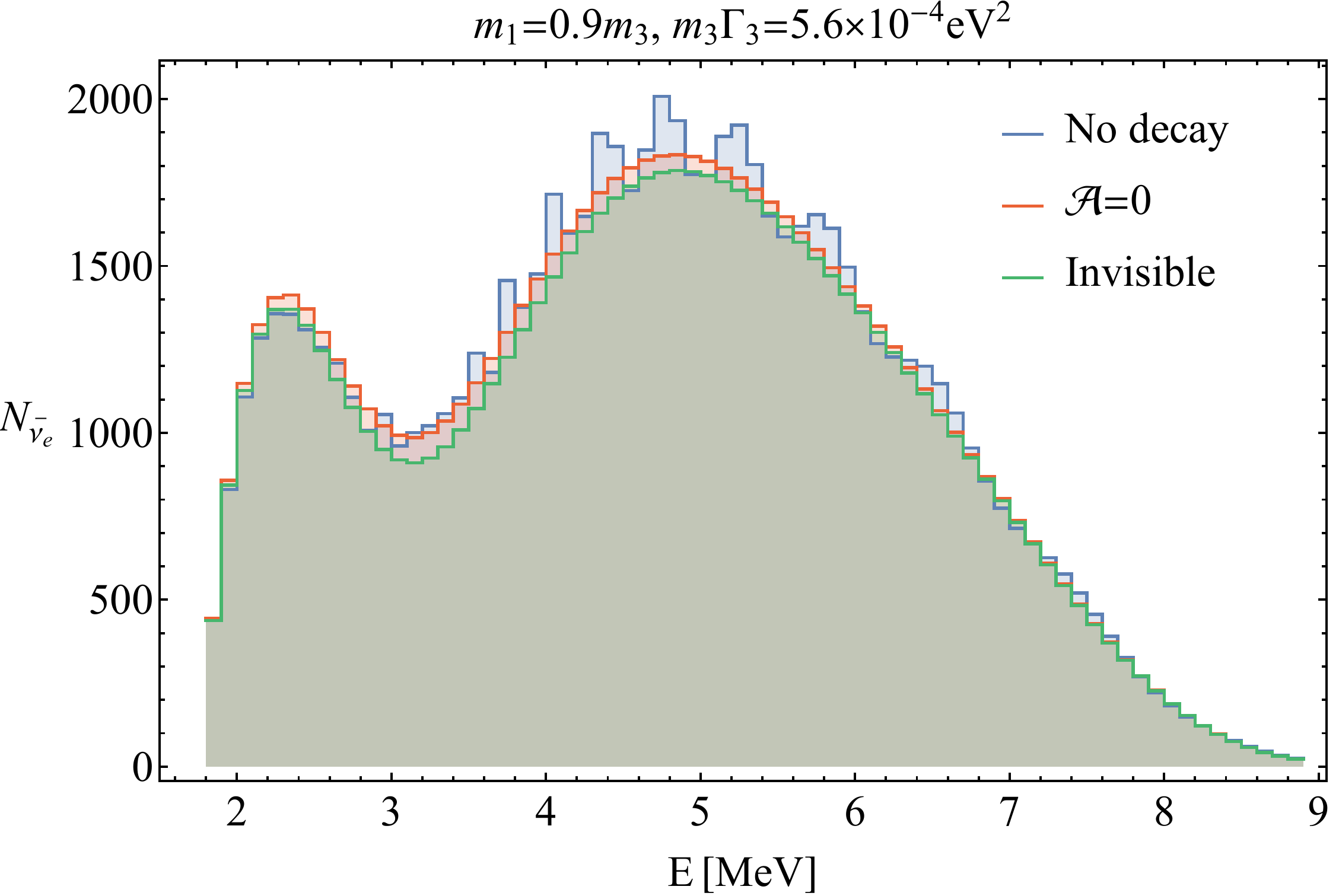}
\caption{Expected number of $\overline{\nu}_e$-mediated events after two years of JUNO data-taking (in energy bins of width 0.1~MeV), for different values of $m_h\Gamma_h$ and the ratio of the daughter-to-parent neutrino masses, assuming the neutrino decay is mediated by Eq.~(\ref{eqn:L}) with $\A=0$, for the decays $\overline{\nu}_2\rightarrow\overline{\nu}_1+\varphi$ (left) and $\overline{\nu}_3\rightarrow\overline{\nu}_1+\varphi$ (right) and the corresponding invisible decays. }
\label{fig:NVisA0}
\end{figure}

When $\A=0$ and $m_l=0$, all decay daughters are invisible. In this case, the expected event distributions are those depicted in Fig.~\ref{fig:Ninv}. For massive daughters, however, a visible component emerges and, as depicted in Fig.~\ref{fig:A0}, the visible decay is expected to surpass the invisible one as $m_l\to m_h$. Fig.\,\ref{fig:NVisA0} depicts the expected number of events at JUNO  after 2 years of data taking where events are binned into energy bins of width $0.1$~MeV  for $\overline{\nu}_2\rightarrow\overline{\nu}_1+\varphi$ (left-hand panel) and  $\overline{\nu}_3\rightarrow\overline{\nu}_1+\varphi$ (right-hand panel), for $m_l=0.9 m_h$. 

\subsection{JUNO Sensitivity}
\label{sec:JUNO_res}

In order to estimate the sensitivity of JUNO to neutrino decay, we simulate 2 years of JUNO data assuming all neutrino mass-eigenstates are stable and use the simulated data to test the hypothesis that the decay $\bar{\nu}_2\to \bar{\nu}_1+\varphi$ is mediated by Eq.~(\ref{eqn:L}) for different values of $\A$ and $\B$. Note that our main aim in this section is to demonstrate the ability of JUNO to distinguish between the two models, given by  $\A=0$, and $\B=0$ in Eq.~(\ref{eqn:L}). Since we are not interested in carefully quantifying the sensitivity of JUNO to neutrino decay, we simplify our analysis by considering the energy resolution function to be a delta function, and assume $100\%$ efficiency of the detector. We concentrate on the decay $\bar{\nu}_2\to \bar{\nu}_1+\varphi$ instead of $\bar{\nu}_3\to \bar{\nu}_1+\varphi$ for convenience, given that its effects are much more pronounced, as discussed earlier. 

\begin{figure}[htbp]
\includegraphics[width=0.48\textwidth]{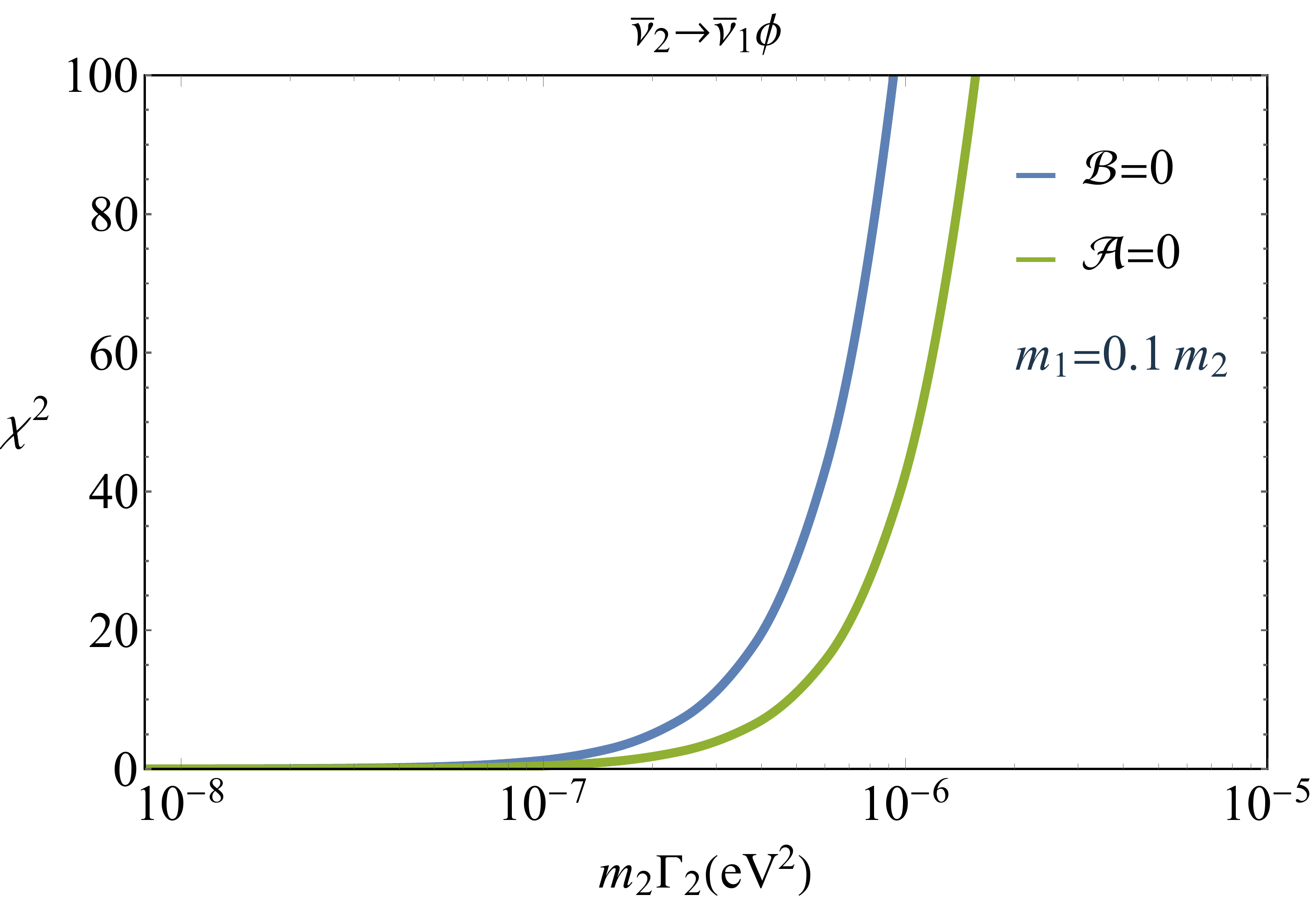}\,\quad\includegraphics[width=0.48\textwidth]{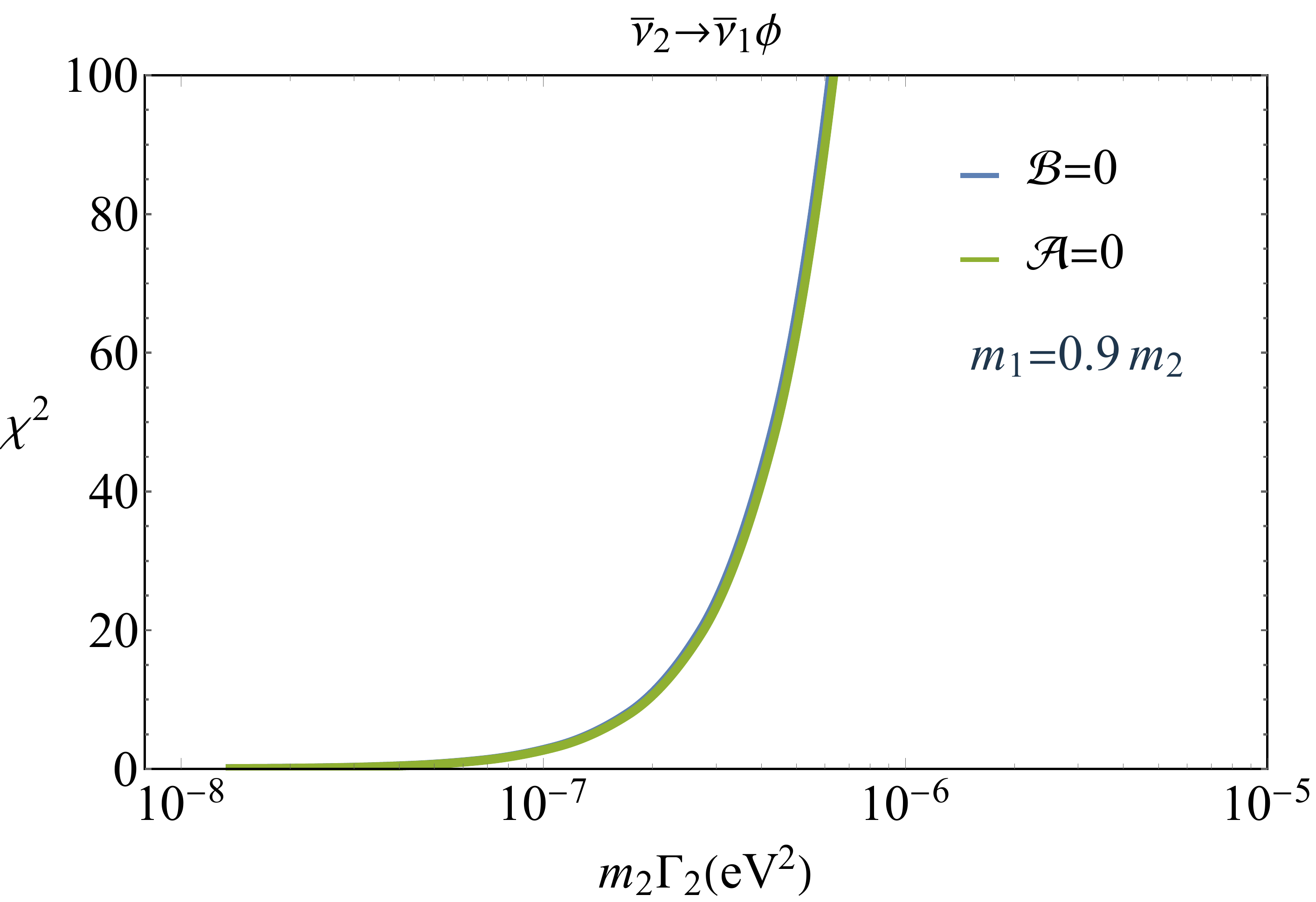}
\caption{$\chi^2$ for the hypothetical decay $\overline{\nu}_2\rightarrow\overline{\nu}_1+\varphi$ as a function of $m_2\Gamma_2$, for different fixed values of $\A$, $\B$, and $m_1/m_2$, assuming two years of JUNO data are consistent with stable neutrinos.}
\label{fig:Chisq_NoDecVsDec}
\end{figure}

Fig.\,\ref{fig:Chisq_NoDecVsDec} depicts $\chi^2$ as a function of $m_2\Gamma_2$ for different fixed values of $\A$, $\B$, and $m_1/m_2$. Here, the simulated data are consistent with stable neutrinos ($\Gamma_2=0$). We assume the oscillation parameters are perfectly known and equal to the values listed in Eq.~(\ref{eq:osc_param}). When the hypothetical value of $m_2\Gamma_2$ is large enough, the decay effects are significant enough that JUNO can distinguish the stable versus unstable $\bar{\nu}_2$ scenarios.

When $\B=0$, independent from $m_1$, the daughter antineutrinos are always visible. Instead, when $m_1\ll m_2$ and $\A=0$, the decay-daughters are mostly invisible. As $m_1\to m_2$, in the $\A=0$ case the fraction of visible daughters approaches one and, in this limit, the $\A=0$ and $\B=0$ case have almost the same signature. This is captured in Fig.\,\ref{fig:Chisq_NoDecVsDec}(right), when $m_1=0.9 m_2$ and the two $\chi^2$-curves virtually lay on top of one another. In the case $m_1=0.1 m_2$, the two models lead to different effects in JUNO. For a fixed value of $m_2\Gamma_2$, JUNO is more sensitive to the $\B=0$ case. In the $\B=0$ case, the visible daughters lead to an enhancement in the expected number of events at JUNO which is reflected in the $\chi^2$ function.

\begin{figure}[htbp]
\includegraphics[width=0.48\textwidth]{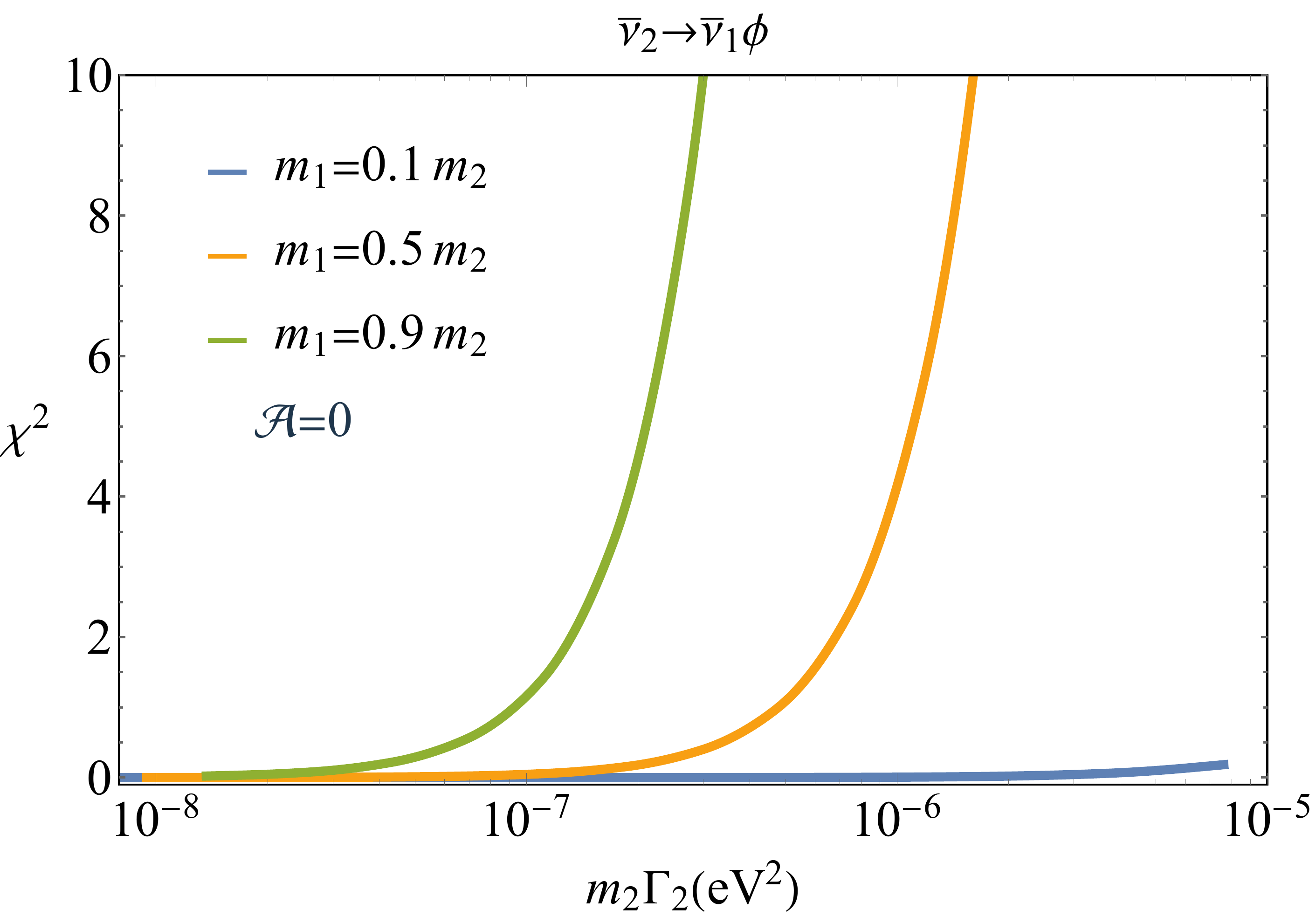}
\includegraphics[width=0.48\textwidth]{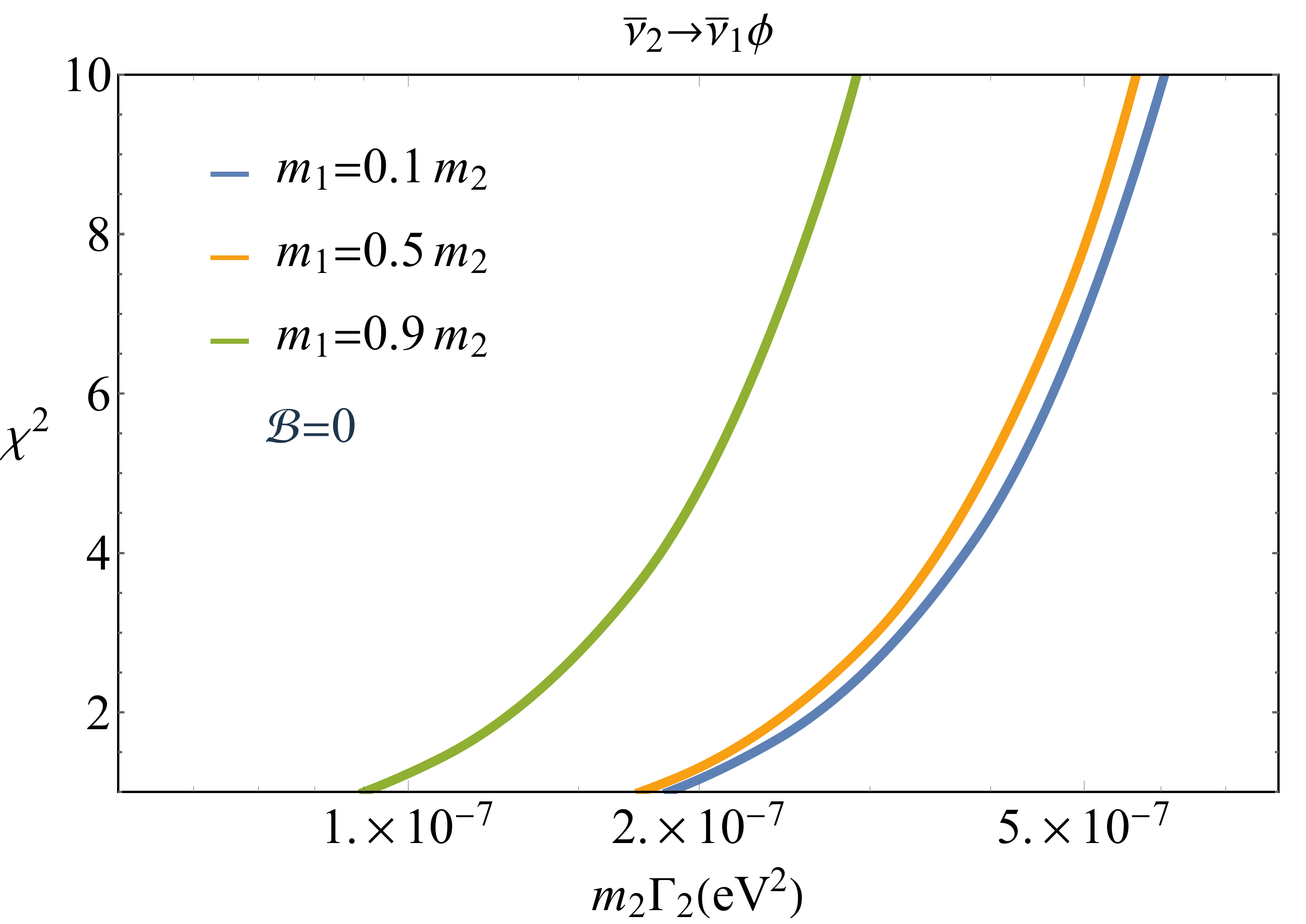}
\caption{Difference in the minimum value of $\chi^2$ when fitting two years of JUNO data consistent with the invisible decay of $\overline{\nu}_2$ to the hypothesis that the decay $\overline{\nu}_2  \to \overline{\nu}_1 +\varphi$, potentially visible, is mediated by Eq.~(\ref{eqn:L}) with either $\A=0$ (left) or $\B=0$ (right), as a function of the input value of $m_2\Gamma_2$, for different values of $m_1/m_2$. }
\label{fig:Chisq_WrongHel}
\end{figure}

In order to further investigate the effects of visible daughters, we simulate 2 years of JUNO data assuming they are consistent with the invisible decay of the $\bar{\nu}_2$ for different values of $m_2\Gamma_2$.  Fig.\,\ref{fig:Chisq_WrongHel} depicts, for two different values of the input value of $m_1/m_2$ and as a function of the input value of $m_2\Gamma_2$, the difference between the minimum value of $\chi^2$ for the invisible decay hypothesis and the assumption that the decay is mediated by Eq.~(\ref{eqn:L}) with $\A=0$  (left-hand panel) or $\B=0$ (right-hand panel). When the input value of $m_2\Gamma_2$ is small, decay effects are negligible and hence it is impossible to distinguish invisible decays from the $\A=0$ or $\B=0$ scenarios, independent from the value of $m_1/m_2$. Hence all curves approach zero as $m_2\Gamma_2\to 0$. For large enough input values of $m_2\Gamma_2$, the fact that some fraction of the decays mediated by the $\A=0$ or $\B=0$ scenarios are visible leads one to disfavor the visible-decay scenario. As expected, when $m_1/m_2$ is small, the majority of the decays in the $\A=0$ scenario are invisible and, quantitatively, one cannot distinguish the two hypothesis for any value of $m_2\Gamma_2$. On the other hand, as $m_1\to m_2$, the fraction of visible decays in the $\A=0$ scenario is high and the two decay hypothesis can be statistically distinguished as long as the ``real'' value of $m_2\Gamma_2$ is large enough (i.e., when one can distinguish the decay hypotheses from the stable-neutrinos hypothesis). In the $\B=0$ case, the visible component is present for all values of $m_1/m_2$ so the decay daughters are always visible, and hence sensitivity to the daughter mass becomes important only when $m_2$ and $m_1$ are degenerate. This leads to the relative shape difference for the $\B=0$ case between the two mass ratios.

\begin{figure}[htbp]
\includegraphics[width=0.48\textwidth]{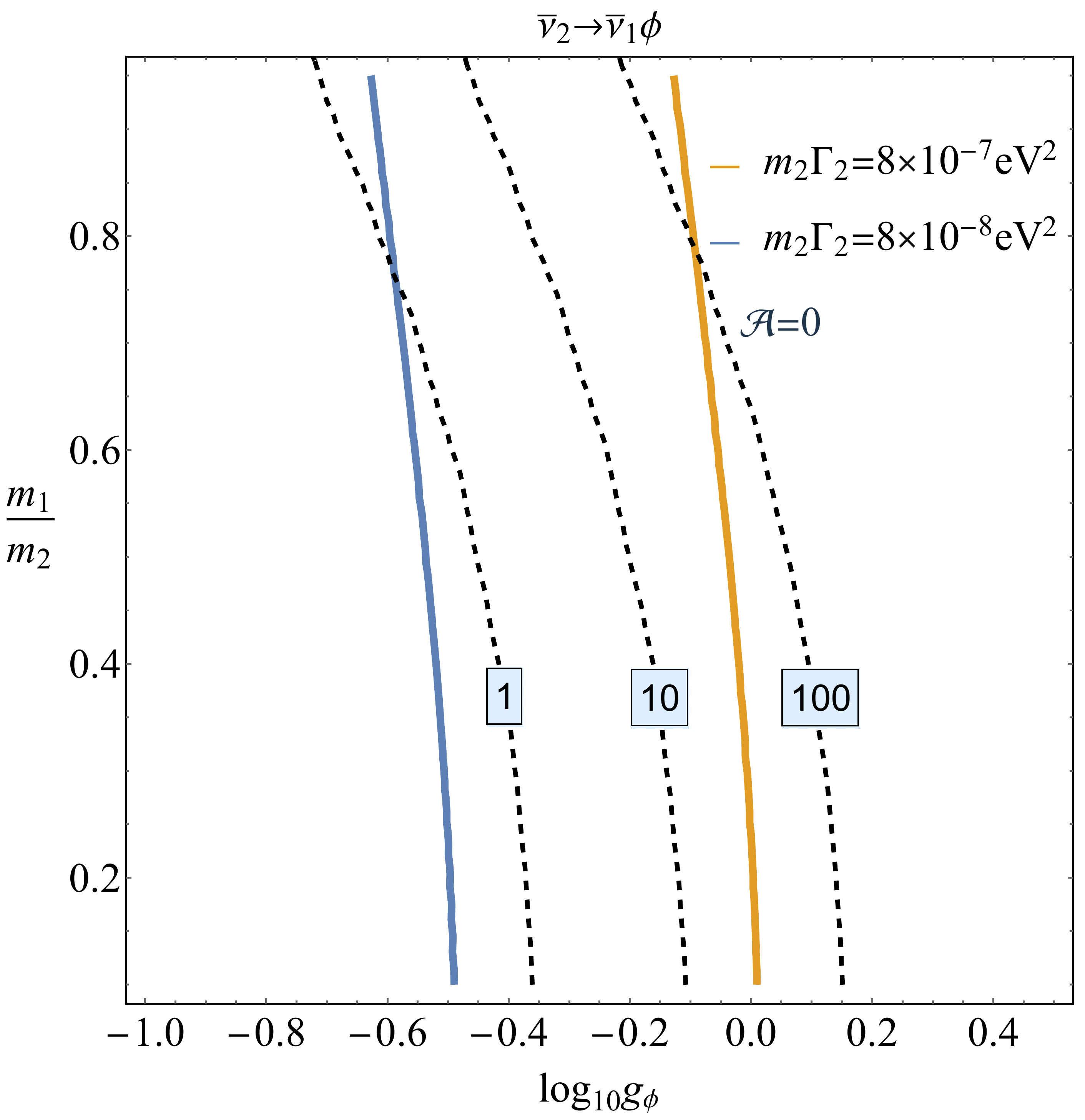}\,\quad\includegraphics[width=0.48\textwidth]{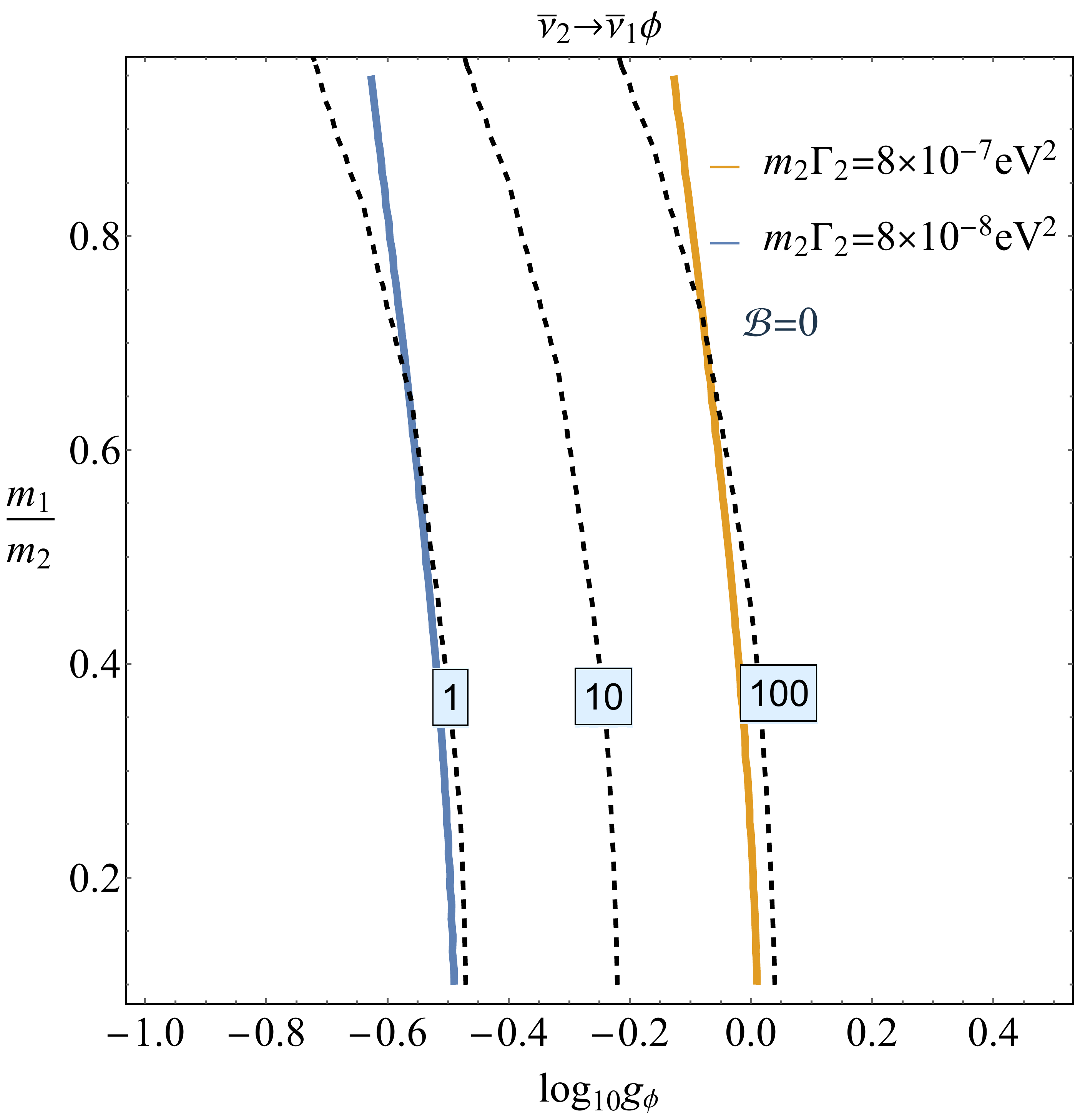}
\caption{Constant $\chi^2=1,10,100$ contours (dashed lines) in the $g_{\varphi}\times m_1/m_2$-plane obtained when fitting two years of simulated JUNO data consistent with stable neutrinos to the hypothesis that the decay $\overline{\nu}_2  \to \overline{\nu}_1 +\varphi$, potentially visible, is mediated by Eq.~(\ref{eqn:L}) with either $\A=0$ (left) or $\B=0$ (right). The solid lines are associated to  constant values of $m_2\Gamma_2$.  }
\label{fig:Chisq_2D}
\end{figure}

Fig.\,\ref{fig:Chisq_2D} depicts the sensitivity of 2 years of JUNO data to neutrino decay assuming it is mediated by the $\A=0$ (left-hand panel) or the $\B=0$ (right-hand panel) models, as a function of $g_{\varphi}$ and the ratio of parent to daughter masses, $m_1/m_2$. In more detail, we simulate data consistent with stable neutrinos and compute the region of parameter space defined by $\chi^2=1,10,100$. For comparative purposes, we depict curves associated to fixed values of $m_2\Gamma_2$.  It is clear that, in both scenarios, JUNO is sensitive to more than $m_2\Gamma_2$ and that the sensitivity improves, relatively speaking, as $m_1\to m_2$. JUNO is more sensitive to the $\B=0$ hypothesis relative to the $\A=0$, especially when $m_1\ll m_2$. This has been discussed before. In the $\B=0$ case, visible daughters are present for all values of $m_1/m_2$ while they are only relevant in the $\A=0$ case when $m_1/m_2$ is not too small. Instead, the sensitivy to both hypotheses is the same in the limit $m_1/m_2\to 1$.

\section{Neutrino three-body decay}
\label{sec:3body}
\setcounter{equation}{0}

In this section, we calculate the differential partial width associated to the three-body decays of neutrinos into lighter neutrinos and antineutrinos. Here, we will focus on the decay $\nu_3\rightarrow\nu_2+\nu_1+\overline{\nu}_1$. This process only occurs if the neutrino mass ordering is normal. It can, however, be directly related to, for example, $\nu_2\to\nu_1+\nu_3+\overline{\nu}_3$, which only occurs if the neutrino mass ordering is inverted. As discussed in Sec.~\ref{sec:models}, we will consider that the neutrino decay is mediated by the ``all-singlets'' dimension-six effective Lagrangian, Eq.~(\ref{eq:3b}). We rewrite it here, labelling the different neutrino states and expressing them as four-component Dirac fields:
\begin{equation}
\label{eqn:L3}
\mathcal{L_{\rm 3-body}}= -G_\nu\,\left(\overline{\nu}_2\gamma^\mu\mathbb{P_R}\nu_3\right)\left(\overline{\nu}_1\gamma_\mu\mathbb{P_R}\nu_1\right)\,.
\end{equation}
We define $G_{\nu}\equiv 1/\Lambda_{\nu}^2$. The three-body decay channel of choice is the ``cleanest'' one since it avoids complications arising from identical particles in the final state.

When all the daughter neutrinos are massless, the daughters of the decay are invisible in any reference frame, i.e., all daughter neutrinos have right-handed helicity and all daughter antineutrinos have left-handed helicity, independent from the helicity of the parent and their other kinematical properties.  However, if any of the daughter neutrinos is massive, based on the results we presented in Sec.~\ref{sec:2body}, we expect some of the daughter neutrinos to be left-handed and some of the daughter antineutrinos to be right-handed, even in the lab frame. The fraction of visible daughters should depend on the helicity of the parent and the masses of the daughters relative to that of the parent.

The amplitude for $\nu_3^{h_3}\rightarrow\nu_2^{h_2}+\nu_1^{h_1}+\overline{\nu}_1^{h_{\bar{1}}}$ is
\begin{equation}
\label{eqn:M}
\left|\mathcal{M}\right|= G_\nu\,\left(\overline{u}_2\gamma^\mu\mathbb{P_R}u_3\right)\left(\overline{u}_1\gamma_\mu\mathbb{P_R}v_{\bar{1}}\right)\,,
\end{equation}
and the magnitude of the amplitude squared is
\begin{eqnarray}
\label{eqn:Msquared}
   \left|\mathcal{M}\right|^2= G_{\nu}^2\left[h_1m_1(h_2m_2S_1\cdot S_2+p_2\cdot S_1)+h_2m_2p_1\cdot S_2+p_1\cdot p_2\right]\cross\nonumber\\
   \left[-h_{\bar{1}} m_{\bar{1}} (h_3 m_3 S_3\cdot S_{\bar{1}}+p_3\cdot S_{\bar{1}})+h_3 m_3 p_{\bar{1}}\cdot S_3+p_3\cdot p_{\bar{1}}\right]\,,
\end{eqnarray}
where 
$p_k,h_k,S_k=(|\mathbf{p_k}|,E_k\hat{p}_k)/m_k$ are, respectively, the four-momenta, helicities, and polarization four-vectors ($S_k^2=-1$ and $S_k\cdot p_k=0$), for $k=1,2,3,\bar{1}$ ($\bar{1}$ indicates the antineutrino). In deriving Eq.~(\ref{eqn:Msquared}), we made use of the following identities for polarized spinors
\begin{eqnarray}
\label{eqn:trick}
   u_\beta(p_k,s_k)\overline{u}_\alpha(p_k,s_k)=\frac{1}{2}[(\not\! p_k+m_k)(1+h_k\gamma_5\not\! S_k)]_{\beta\alpha},\nonumber\\
   v_\beta(p_k,s_k)\overline{v}_\alpha(p_k,s_k)=\frac{1}{2}[(\not\! p_k-m_k)(1+h_k\gamma_5\not\! S_k)]_{\beta\alpha}\,.
\end{eqnarray}
The differential decay width is
\begin{equation}
\label{eqn:dGamma}
d\Gamma=\frac{1}{(2\pi)^5}\frac{1}{2E_3}\left|\mathcal{M}\right|^2d\phi_3\,,
\end{equation}
where
\begin{equation}
\label{eqn:dPhi}
d\phi_3=\frac{d^3p_1}{2E_1}\frac{d^3p_2}{2E_2}\frac{d^3p_{\bar{1}}}{2E_{\bar{1}}}\delta^{(4)}(p_3-p_1-p_2-p_{\bar{1}})\,,
\end{equation}
is the three-body phase space.

In order to render the discussion and results more transparent, we will concentrate on the case $m_1=0$. In this case, the helicities of $\nu_1$ and $\bar{\nu}_1$ are, independent from the reference frame, always $h_1=+1$, $h_{\bar{1}}=-1$ and these states are always invisible. This implies that, when it comes to phenomenology, we will mostly be interested in the differential partial width as a function of the energy and direction of the daughter $\nu_2$. With this in mind, taking advantage of the assumption $m_1=0$, we integrate over the ``invisible'' part of the phase space using the following identity:
\begin{equation}
\label{eqn:Michel}
\int \frac{d^3p_1}{2E_1}\frac{d^3p_{\bar{1}}}{2E_{\bar{1}}}\delta^{(4)}(Q-p_1-p_{\bar{1}})p_1^\alpha p_{\bar{1}}^\beta=\frac{\pi}{24}\left(g^{\alpha\beta}Q^2+2Q^\alpha Q^\beta\right)\,,
\end{equation}
where $Q=p_3-p_2$. 
In the next two subsections, we discuss the $\nu_3$ decay, first in its rest frame, then in the laboratory frame.

Before proceeding, we highlight that since we are interested in the case where the lightest neutrino is massless ($m_1=0$ in the case of the normal mass ordering), the values of the nonzero neutrino masses can be determined from the mass-squared differences. For the normal mass ordering, using the values in Eq.~(\ref{eq:osc_param}),
\begin{equation}
\label{eqn:NOval}
m_2 = 0.0087~{\rm eV},~~~~m_3= 0.0505~{\rm eV},~~~~m_2/m_3 = 0.175.
\end{equation}
while for the inverted mass ordering, when the relevant decay is $\nu_2\to\nu_1\nu_3\bar{\nu}_3$ and $m_3=0$,
\begin{equation}
\label{eqn:IOval}
m_1 = 0.0497~{\rm eV},~~~~m_2 = 0.0505~{\rm eV},~~~~m_1/m_2 = 0.985.
\end{equation}
Regardless, in the discussions below, we will treat the parent-neutrino ($m_3$) and daughter-neutrino mass ($m_2$) as unconstrained parameters. 

\subsection{Parent neutrino rest frame}

In the rest frame of the parent neutrino, there is no preferred direction other than its polarization, which we use to define the $z$-axis. Therefore, 
\begin{eqnarray}
    p_3^*&=&(m_3,0,0,0),\nonumber \\
   p_2^* &=& (E_2^*,|\mathbf{p_2^*}|\sin\theta_2^*,0,|\mathbf{p_2^*}|\cos\theta_2^*)\,.
\end{eqnarray}
We use starred variables to refer to kinematical variables in the rest frame.
Plugging these into Eq.~(\ref{eqn:dGamma}), and using Eq.~(\ref{eqn:Michel}),
\begin{eqnarray}
\label{eqn:d2GammaRest}
   \frac{d^2\Gamma^*}{dE_2^* d\cos\theta_2^*}=\mathcal{C}(E_2^*,m_3,m_2,h_2^*) + \mathcal{D}(E_2^*,m_3,m_2,h_2^*,h_3^*) \cos\theta_2^*\,,
\end{eqnarray}
where $\mathcal{C}, \mathcal{D}$ are functions arising out of the partial phase space integration; these functions are listed in Appendix A. The $\mathcal{D}$ coefficient, which governs the nontrivial $\cos\theta_2^*$ dependency, depends on the parent polarization\footnote{As discussed earlier, in the parent rest-frame, the helicity is not well defined. Instead, we use $h_3^*$ to indicate the parent polarization. $h_3^*=1$ indicates spin ``up'' while $h_3^*=-1$ indicates spin ``down.''} while the $\mathcal{C}$ coefficient, which governs the $\cos\theta_2^*$ independent term, does not. The dependence on $h_3^*$ vanishes upon integration over $\cos\theta_2^*$. This is easy to understand and a consequence of the fact that, in the rest frame of the parent, there is no preferred direction other than its polarization. 

\begin{figure}[!t]
\includegraphics[width=.75\textwidth]{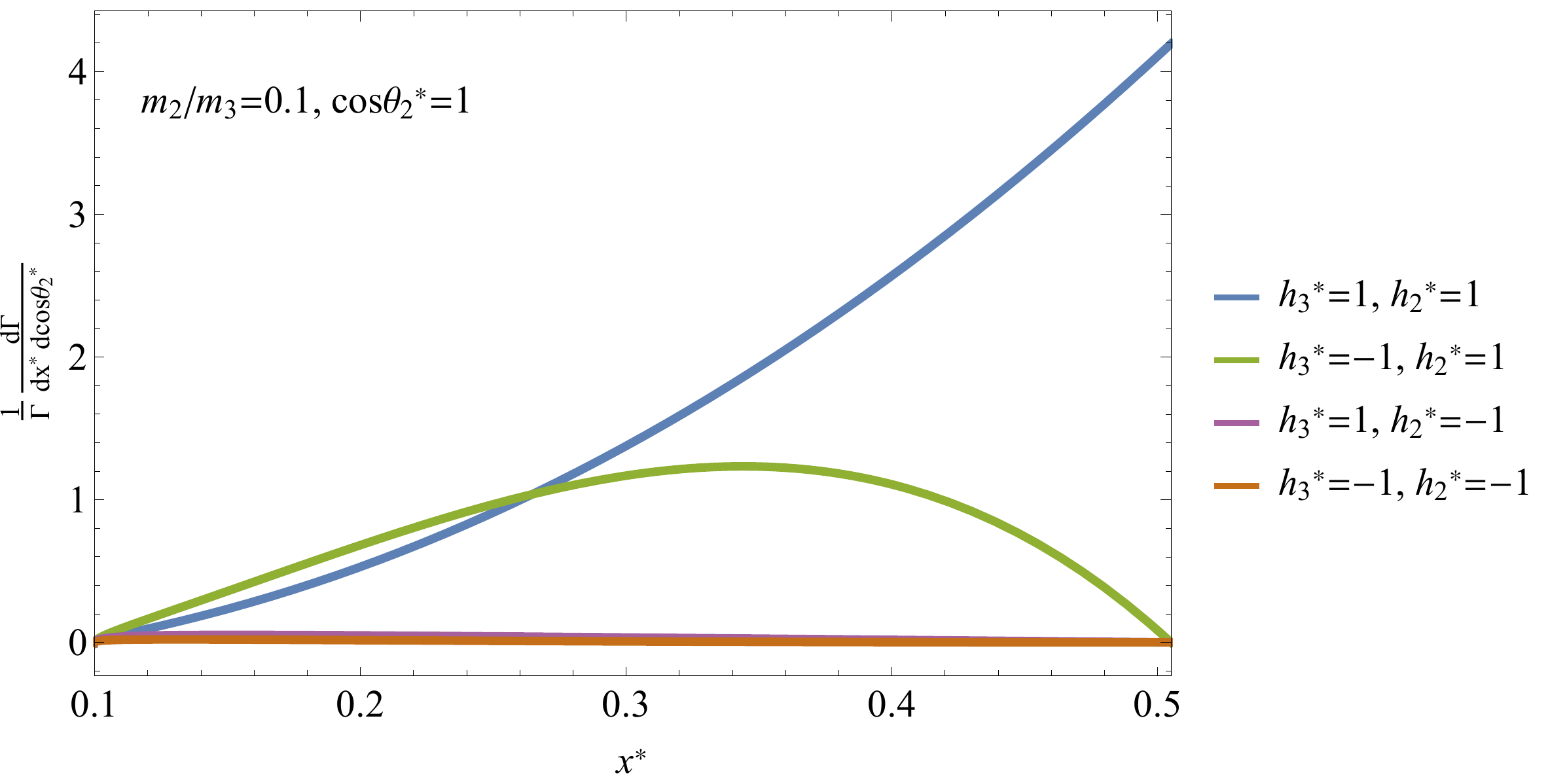}
\caption{Double-differential decay width normalized to the total width as a function of $x^*=E_2^*/m_3$ for $\nu_3\to\nu_2\nu_1\overline{\nu}_1$ in the parent rest frame when $\cos\theta_2^*=1$, and $m_2=0.1~m_3$, for different values of the parent polarization and daughter helicity ($h_3^*$ and $h_2^*$, respectively).}
\label{fig:cos1}
\end{figure}

\begin{figure}[!t]
\includegraphics[width=0.7\textwidth]{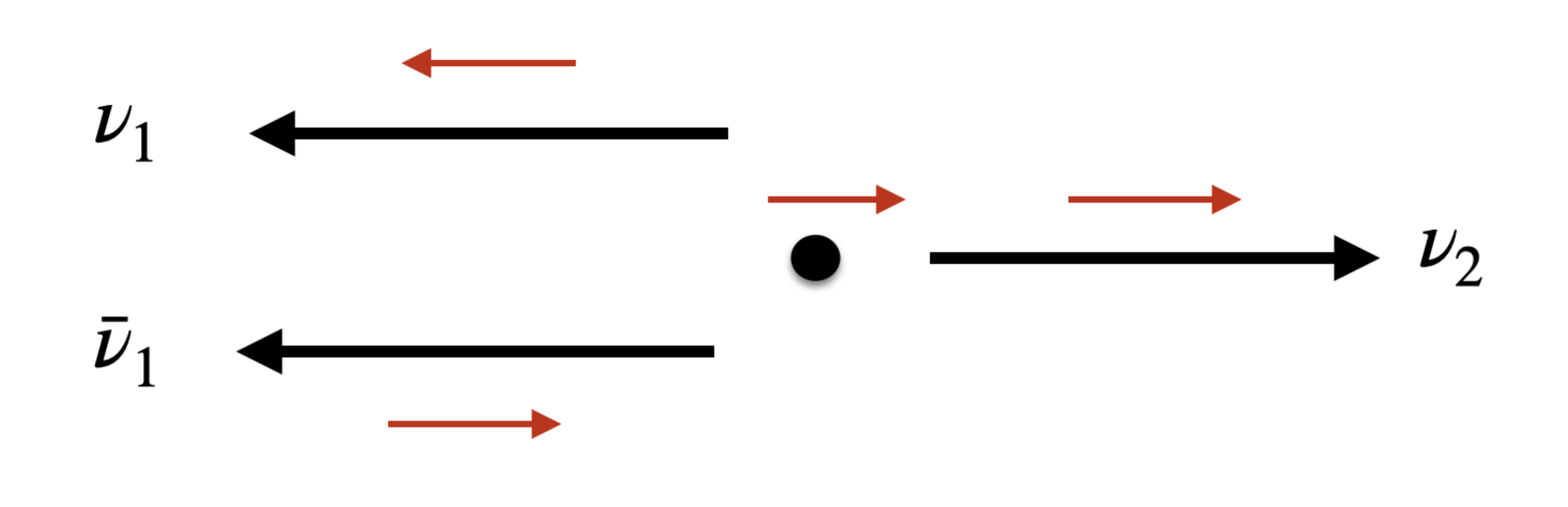}
\caption{Illustration of the daught-helicity configuration in $\nu_3\to\nu_2\nu_1\overline{\nu}_1$ in the parent rest-frame when $\cos\theta^*_2=1$ and $h^*_2=1$. Since, by assumption, $m_1$ is zero, $h^*_1=1$ and $h^*_{\bar{1}}=-1$ for all kinematical configurations in all reference frames. }
\label{fig:kinconfpos}
\end{figure}

Fig.~\ref{fig:cos1} depicts the double-differential decay width when $\cos\theta_2^*=1$, and $m_2=0.1~m_3$, for different values of the parent and daughter helicities ($h_3^*$ and $h_2^*$, respectively). Some features of the case where $h_2^*=1$ -- chirality-favored keeping in mind the right-chiral nature of the interaction that governs the decay -- can be qualitatively understood. For $\cos\theta_2^*=1$, angular and linear momentum conservation allow the daughters to be emitted such that the $\nu_2$ reaches its maximum allowed energy $(E_2^*)_{\rm max}\sim m_3/2$ (when $m_2\lll m_3$). In this case, the $\nu_1$ and $\overline{\nu}_1$ are emitted ``backwards'' and share the remaining energy:
\begin{eqnarray}
    p_2^*&=&(m_3/2,0,0,m_3/2),\nonumber \\
    p_1^* &=& p_{\bar{1}}^*=(m_3/4,0,0,-m_3/4).
\end{eqnarray}
This kinematic configuration is illustrated in Fig.\,\ref{fig:kinconfpos}. In this special case, the combined $\nu_1$--$\overline{\nu}_1$ behaves like a massless spin-0 object, and, because of angular momentum conservation, the daughter $\nu_2$ must have the same spin orientation as the parent neutrino $\nu_3$. Note that these linear arguments in the rest frame are of special interest since they could extend to the lab frame. For a $\hat{z}$ boost, there would be a trivial correspondence between the helicities of each particles in the two frames. This behavior is manifest in Fig.\,\ref{fig:cos1} where, at $(E_2^*)_{\rm max}$, the helicity preserving channel peaks, while the helicity flipping channel vanishes. The chirality-disfavored $h_2^*=-1$ channel, instead, is very suppressed independent from $h_3^*$. This is a consequence of the fact that, in Fig.~\ref{fig:cos1}, $m_2\ll m_3$.


\begin{figure}[!t]
	\includegraphics[width=0.7\textwidth]{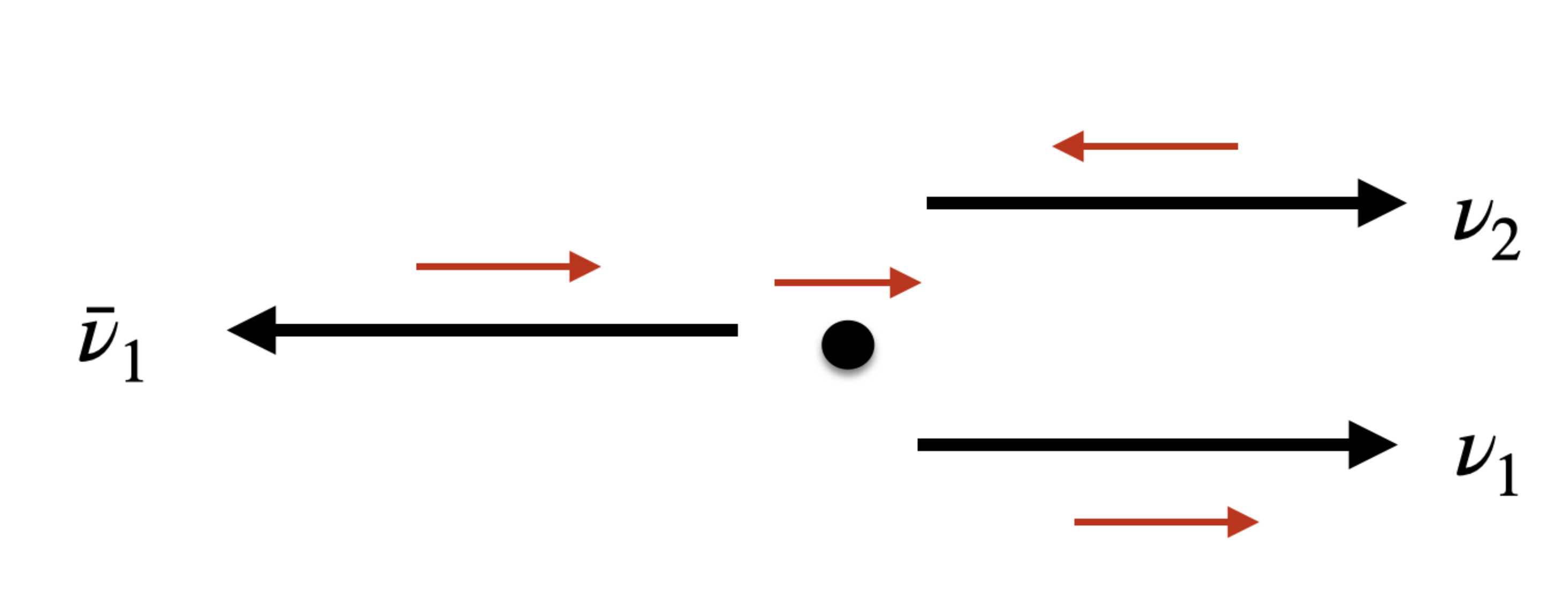}
	\caption{Illustration of the daught-helicity configuration in $\nu_3\to\nu_2\nu_1\overline{\nu}_1$ in the parent rest-frame when $\cos\theta^*_2=1$ and $h^*_2=-1$. Since, by assumption, $m_1$ is zero, $h^*_1=1$ and $h^*_{\bar{1}}=-1$ for all kinematical configurations in all reference frames. }
	\label{fig:kinconfneg}
\end{figure}

\begin{figure}[!t]
	\includegraphics[width=.7\textwidth]{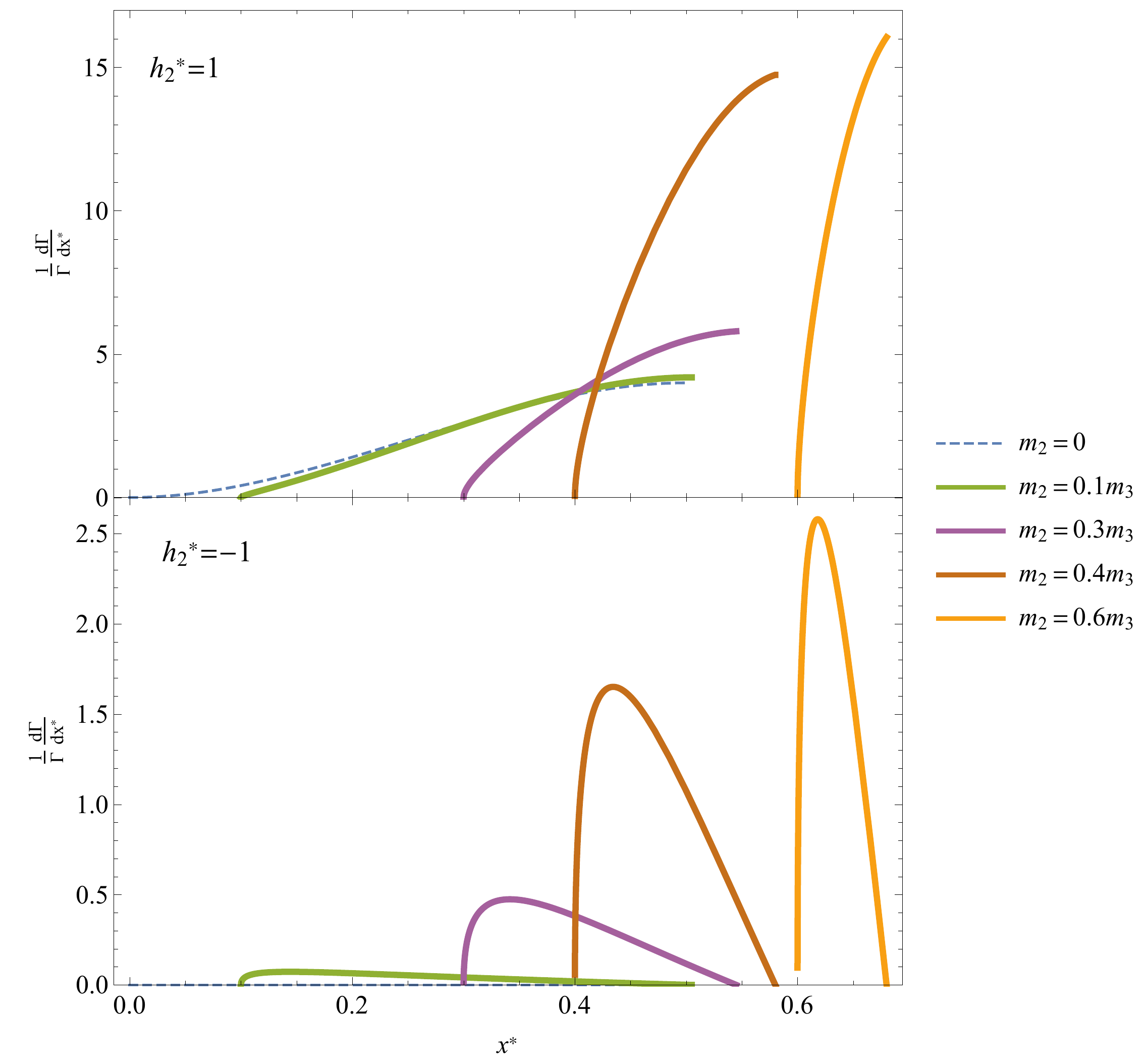}
	\caption{Differential decay width normalized to the total width as a function of $x^*=E^*_2/m_3$, for $\nu_3\rightarrow \nu_2+\nu_1+\overline{\nu}_1$ assuming $m_1=0$ and different values of $m_2$ and the $\nu_2$ helicity $h_2^*$, in the $\nu_3$ rest frame. Since $m_1=0$, given the nature of the interaction that governs the decay (see Eq.~(\ref{eqn:L3})), $h^*_1=1$ and $h^*_{\bar{1}}=-1$.}
	\label{fig:rest}
\end{figure}

There is some added complexity in the three-body decay relative to the two-body one. In the rest frame, the amplitude squared for the three-body decay is
\begin{equation}
\label{eqn:matrest}
|\mathcal{M}_{h_3^*h_2^*}|^2 = 4|G_\nu|^2m_3E_1^*E_{\bar{1}}^*\left(1-h_3^*\cos\theta_{\bar{1}}^*\right) \, \times
\left\{
\begin{array}{l@{\quad}l}
\left(E_2^*+\sqrt{E_2^{*2}-m_2^2}\right)\left(1-\cos\left(\theta_1^*-\theta_2^*\right)\right)\,, & h_2^*=1 \\
\left(E_2^*-\sqrt{E_2^{*2}-m_2^2}\right)\left(1+\cos\left(\theta_1^*-\theta_2^*\right)\right)\,, & h_2^*=-1
\end{array} \right. \,,
\end{equation}
where $E^*_1,E^*_{\bar{1}}$ are, respectively, the energy of $\nu_1,\bar{\nu}_1$ and $\theta^*_{\bar{1}}$ is the angle defined by the momentum of $\bar{\nu}_1$ relative to the $z$-axis. If $h_2^*=-1$, a non-vanishing contribution requires $m_2\neq 0$, $h_3^*\neq \cos\theta_{\bar{1}}^*$, and $\cos\theta_1^*\neq -\cos\theta_2^*$. The kinematical configuration in Fig.~\ref{fig:kinconfpos} clearly violates the third condition, and spin conservation prevents the second one. The maximum contribution corresponds to Fig.~\ref{fig:kinconfneg}
where, in this case, the combined $\nu_1$--$\overline{\nu}_1$ behaves like a massless spin-1 object. The $\nu_1$ and $\nu_2$ must share some energy which constraints $E_2^*$ to be smaller than its theoretical kinematical maximum. This behavior will be visible in Fig.\,\ref{fig:rest}, when we allow for larger values of $m_2/m_3$. To summarize, very roughly speaking, in the $h_2^*=1$ case, the combined $\nu_1$--$\overline{\nu}_1$ wants to behave like a spin-0 object, while in the $h_2^*=-1$, it wants to behave as a spin-1 object.

Integrating over $\cos\theta_2^*$, Fig.\,\ref{fig:rest} depicts the differential decay width as a function of the daughter neutrino energy $E_2^*$ for different values of $m_2$. As discussed earlier, in these distributions do not depend on the polarization state of the parent. For  $m_2\ll m_3$ the probability of emitting a left-handed $\nu_2$ ($h_2^*=-1$) is very small, as is expected from the chiral nature of the interaction. However, for larger values of $m_2/m_3$ the significance of the $h_2^*=-1$ channel grows.

Finally, integrating over $E_2^*\in[m_2,(m_3^2+m_2^2)/2m_3]$, we find 
\begin{equation}
\Gamma^*(h_2^*=1)=\frac{G^2_\nu m_3^5}{4608\pi^3}\left[3-32\frac{m_2^2}{m_3^2}+48\frac{m_2^3}{m_3^3}-\frac{m_2^4}{m_3^4}\left(45+36\log \frac{m_2}{m_3}\right)+16\frac{m_2^5}{m_3^5}+12\frac{m_2^6}{m_3^6}-2\frac{m_2^8}{m_3^8}\right]\,,
\end{equation}
and
\begin{equation}
\Gamma^*(h_2^*=-1)=\frac{G^2_\nu m_3^5}{4608\pi^3}\left\{\frac{m_2^2}{m_3^2}\left[8-48\frac{m_2}{m_3}+\frac{m_2^2}{m_3^2}\left(45-36\log \frac{m_2}{m_3}\right)-16\frac{m_2^3}{m_3^3}+12\frac{m_2^4}{m_3^4}-\frac{m_2^6}{m_3^6}\right]\right\}\,.
\end{equation}
As expected, given the right-chiral nature of the interaction mediating the decay, the $h_2^*=-1$ decay mode occurs only when $m_2\neq 0$ and its partial width is suppressed by a factor of order $(m_2/m_3)^2$ relative to the chirality preferred $h_2^*=1$ final state.

\subsection{Laboratory Frame}

Here, we repeat the analysis of the last subsection, this time concentrating on the lab frame, in the limit where the energy of the parent $E_3$ is significantly larger than its mass. Since we are interested in the helicities of the daughter neutrinos, naively boosting the differential decay widths computed in the rest frame of the parent neutrino is insufficient, since the helicities of massive fermions are also reference-frame dependent. We find that it is easiest to compute the differential decay widths directly in the laboratory reference frame.

We define the $z$-axis to coincide with the direction of motion of the parent neutrino. Hence,
\begin{eqnarray}
    p_3&=&(E_3,0,0,|\mathbf{p_3}|),\nonumber \\
   p_2 &=& (E_2,|\mathbf{p_2}|\sin\theta_2,0,|\mathbf{p_2}|\cos\theta_2)\,.
\end{eqnarray}
Using Eq.~(\ref{eqn:Michel}) to integrate over the invisible $\nu_1,\bar{\nu}_1$ part of the phase space,
\begin{eqnarray}
   \frac{d^2\Gamma}{dE_2d\cos\theta_2}=\mathcal{F}_1(E_3,E_2,m_3,m_2,h_2,h_3) + \mathcal{F}_2(E_3,E_2,m_3,m_2,h_2,h_3) \cos\theta_2 +  \mathcal{F}_3(E_3,E_2,m_3,m_2,h_2,h_3) \cos\,2\theta_2\,,\nonumber\\
   \label{eq:2dgamma}
\end{eqnarray}
where $\mathcal{F}_{1,2,3}$, listed in the Appendix A, are functions arising out of phase space integration. In order to compute $d\Gamma/dE_2$, we need to integrate over $\cos\theta_2$. Since in typical experiments of interest neutrinos are ultrarelativistic, the daughter neutrinos produced in the decay are emitted in the forward direction and lie within a very narrow cone around the $z$-axis. In more detail,
\begin{equation}
\cos\theta_2=\left(1+\left(\frac{\mathbf{|p_2^*|}\sin\theta_2^*}{\gamma\left(\mathbf{|p_2^*|}\cos\theta_2^*+\beta E_2^*\right)}\right)^2\right)^{-1/2}\,,
\label{eqn:rela1}
\end{equation}
where the starred-variables refer to the rest-frame parameters while $\beta,\gamma$ are the traditional parameters that define the boost between the laboratory and the $\nu_3$ rest-frame, $\gamma=E_3/m_3$ and $\beta=|\mathbf{p_3}|/E_3$. 
In the ultrarelativistic limit ($\beta\to 1$, $\gamma\to\infty$), $\cos\theta_2\to 1$ for any $\cos\theta_2^*\neq -1$.\footnote{Some care is needed for $\cos\theta_2^*$ values very close to $-1$.} 
\begin{figure}[!t]
	\includegraphics[width=1\textwidth]{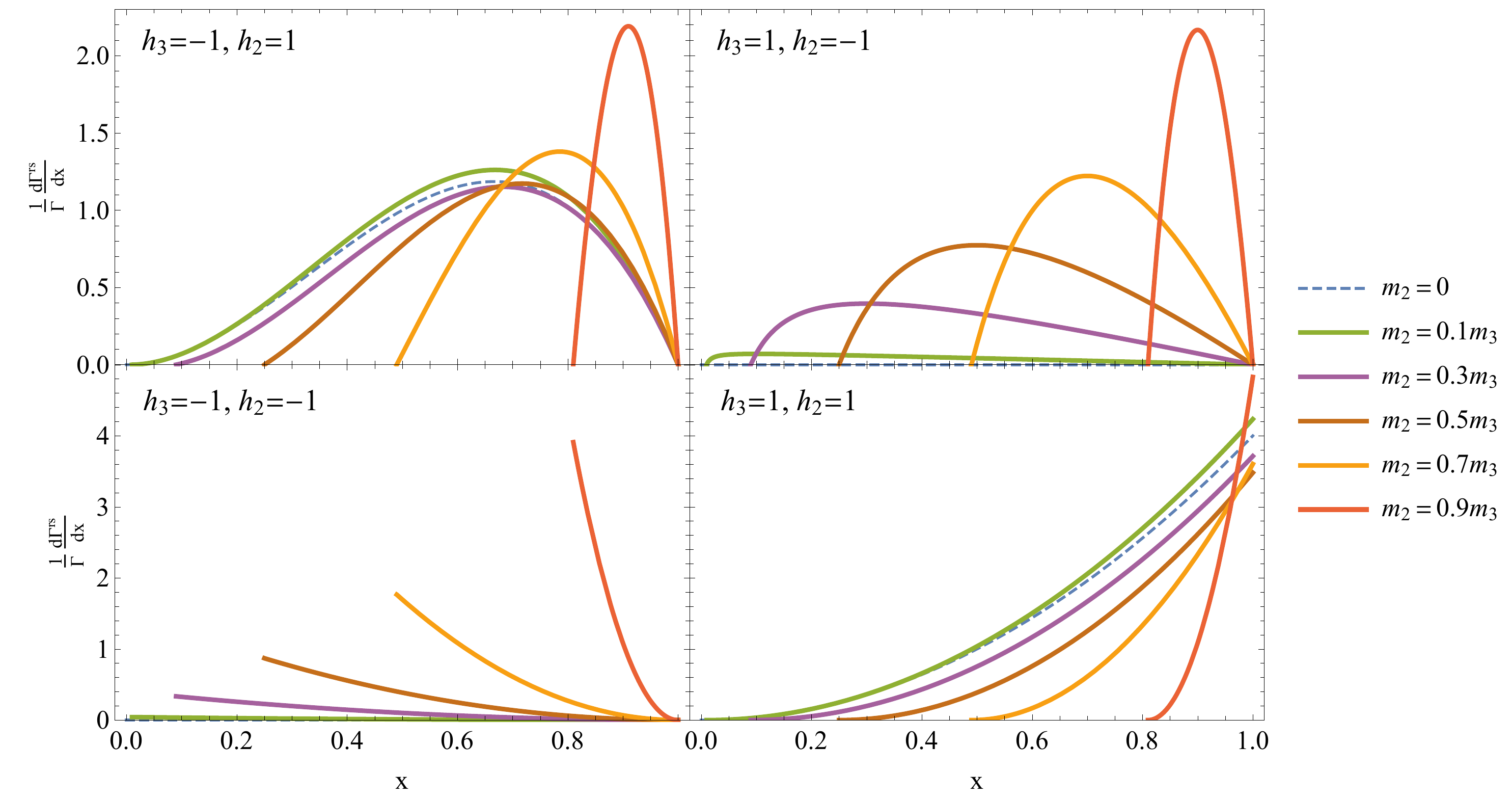}
	\caption{Differential decay width in the lab frame normalized to the total width as a function of $x=E_2/E_3$, for $\nu_3\rightarrow \nu_2\nu_1\overline{\nu}_1$ assuming different values of $m_2/m_3$, $\nu_2$, the $\nu_2$ helicity $h_2$, and the $\nu_3$ helicity $h_3$. Since $m_1=0$ has been assumed, given the nature of the interaction that governs the decay (see Eq.~(\ref{eqn:L3})), $h_1=1$ and $h_{\bar{1}}=-1$. }
	\label{fig:lab}
\end{figure}

Taking advantage of the very forward nature of the daughter neutrinos, we trivially integrate Eq.~(\ref{eq:2dgamma}) over $\cos\theta_2$ by setting it to one.\footnote{This is a relatively standard procedure. It is used, for example, in order to compute the differential neutrino fluxes at neutrino factories, see, for example, Ref.~\cite{Apollonio:2002en}. Some care is needed in order to ensure the value of the lifetime is not corrupted along the way.} The results are depicted in Fig.\,\ref{fig:lab} for different values of $m_2/m_3$. As expected, when $m_2\ll m_3$ only right-handed $\nu_2$ ($h_2=1$) are emitted. 
As $m_2$ increases, the contribution of the chirality-disfavored helicity increases. 

We are especially interested in left-handed parents ($h_3=-1$) since virtually all parent neutrinos that are experimentally accessible are produced via the weak interactions and hence left-handed in the laboratory frame. Moreover, visible daughters, detected via the weak interactions, must also be left-handed ($h_2=-1$) in the laboratory frame. Concentrating on the bottom-left panel in Fig.\,\ref{fig:lab}, we note that as $m_2$ approaches $m_3$, not only is it possible to produce visible daughters when the decay  interaction is right-chiral but there are kinematical regimes where the $h_2=-1$ dominates the differential decay width. As we discussed in the context of two-body decays, the contribution of the chirality-disfavored helicity is proportional to the square of the ratio of the daughter mass to the parent mass. This is true both in the rest frame of the parent and in the lab frame, when the parents (and the daughters) are ultrarelativisitc. 

Some comments on the comparison of the two-body and three-body decays discussed here are in order. Given what is known about neutrino masses, in two-body decay scenarios, the ratio of parent to daughter masses is allowed to vary while in the three-body decay we discussed here, where the lightest neutrino is assumed to be massless, this ratio is fixed, as given by Eq.~(\ref{eqn:NOval}) for NO, and Eq.~(\ref{eqn:IOval}) for IO. In case of IO, the ratio $m_1/m_2$ is very close to one, and hence almost all of the energy of the parent neutrino is transferred to the massive daughter. In such a situation, the two- and three body-decay kinematics, as far as the visible daughter neutrino is concerned, are similar. On the other hand, for NO, a smaller value of the mass ratio would generate a more notable distinction between the energy distribution of the massive daughter for the three-body decay relative to the two-body decay. However, a smaller value of the mass ratio also significantly diminishes the likelihood that visible daughters are produced, as noted in the preceding discussions. This would make it very challenging to differentiate the two decay mechanisms in an upcoming experiment like JUNO. This picture can change, of course, if, in the three-body decays, the lightest daughter neutrino also has a nonzero mass. In this case, one expects different signatures between two-body and three-body decays that can be used to differentiate between the two cases. For example, if all neutrino masses are of the same order of magnitude, it is possible that two of the decay-daughters are visible even if the decay process is mediated by Eq.~(\ref{eqn:L3}). If all daughters of the three-body decay are massive, however, the expression for the decay width cannot be simplified using the ``trick'' advocated in Eq.~(\ref{eqn:Michel}) and hence a different analysis from the one presented here is required. This lies beyond the scope of our current work, and will be pursued in future work.

\section{Conclusion}
\label{sec:conclusion}
\setcounter{equation}{0}

Given everything we have learned about neutrinos in the last 25 years, it is widely anticipated that the two heaviest neutrinos have a finite lifetime. If the weak interactions are the dominant contributions to the neutrino lifetimes, these are expected to be many orders of magnitude longer than the current age of the universe. The presence of new interactions and new very light states, however, can lead to neutrino lifetimes that are orders of magnitude smaller than a second. While it is often the case that these hypothetical new interactions are best constrained by experimental probes outside of neutrino physics, there are scenarios that ``only'' manifest themselves through the neutrino lifetime. 

One such scenario is the existence of new interactions that involve only right-handed neutrino fields and new heavy states, assuming that neutrinos turn out to be Dirac fermions. The four-fermion operator Eq.~(\ref{eq:3b}), for example, involves only gauge-singlet fermions, is unconstrained by all foreseeable laboratory experiments, and, at leading order, only mediates neutrino--neutrino elastic scattering and neutrino decay. 

Here, we computed the kinematical properties of two-body and three-body neutrino decays, assuming the neutrinos are Dirac fermions. We were interested in the helicities of the daughter-neutrinos, especially  when these are produced in interactions involving right-chiral neutrino fields. It is widely known that, in the limit where the daughter-neutrino mass vanishes, the neutrino helicity is dictated by the chirality of the neutrino field, e.g.~right-chiral neutrino fields produce massless right-handed helicity neutrinos. Not surprisingly, when the daughter-neutrino mass is not zero, the right-chiral neutrino fields can produce left-handed helicity neutrinos. We find that this is indeed the case and, more importantly, the probability of producing the chirality-disfavored-helicity daughters is proportional to the ratio of the daughter-mass-squared to the parent-mass-squared both in the parent rest-frame and in the lab frame (as opposed to, for example, the daughter-mass-to-energy ratio). Hence, these effects can be significant even in the laboratory, where neutrinos are ultrarelativistic.

The helicity of the daughter-neutrinos significantly impacts the way in which neutrino decays manifest themselves experimentally. Given that neutrinos are, to an excellent approximation, both ultrarelativistic in the lab frame and produced and detected via the charged-current weak interactions, and given the parity-violating nature of the weak interaction, right-handed-helicity neutrinos and left-handed-helicity antineutrinos are invisible. Left-handed-helicity neutrinos and right-handed-helicity antineutrinos are, on the other hand, potentially visible. Here we also explored whether visible and invisible decays can lead to distinguishable experimental signatures, concentrating on the JUNO experimental setup and two-body neutrino decays, and argued that this is indeed the case.     

Combining these different results, we showed that if the neutrinos are Dirac fermions and they decay via interactions that involve right-handed neutrino fields, some fraction of the decays will contain visible daughters as long as the daughter-neutrino mass is not much smaller than the parent-neutrino mass. This implies that new, ``all singlets'' interactions not only mediate fast neutrino decay that is unconstrained by non-neutrino experiments but also some of the decay products can be detected, as long as the daughter-neutrino masses are not too small relative to the parent mass. The observation of such a decay process would not only reveal new physics but would also carry nontrivial, unique information on the neutrino masses.   

\appendix
\section{Differential Decay Coefficients}
\label{sec:appendix}
\noindent In this Appendix, we provide analytical expressions of the quantities introduced in Eq.~(\ref{eqn:d2GammaRest}), and Eq.~(\ref{eq:2dgamma}).
As discussed, in the rest frame, the differential decay width for the three-body decay can be written as
\begin{eqnarray}
\label{eqn:d2GammaRest_a}
   \frac{d^2\Gamma^*}{dE_2^* d\cos\theta_2^*}=\mathcal{C}(E_2^*,m_3,m_2,h_2^*) + \mathcal{D}(E_2^*,m_3,m_2,h_2^*,h_3^*) \cos\theta_2^*\,.
\end{eqnarray}
The functions $\mathcal{C}, \mathcal{D}$ are given by
\begin{equation}
\begin{split}
\mathcal{C}(E_2^*,m_3,m_2,h_2^*)&=\frac {G_\nu^2}{384\pi^3}\sqrt {E_2^{*2} - 
	m_ 2^2}\left (m_ 3\left( 
3 m_ 3 -4E^*_2 \right)\left (h^*_ 2\sqrt {E_2^{*2} - m_ 2^2} + E^*_2 \right)\right.\\
&\left.+m_ 2^2\left (h^*_ 2\sqrt {E_2^{*2} - m_ 2^2} + 3E^*_2 - 
2 m_ 3 \right) \right)\,,
\end{split}
\end{equation}
and 
\begin{equation}
\begin{split}
\mathcal{D}(E_2^*,m_3,m_2,h_2^*,h_3^*)&=\frac {G_\nu^2}{384\pi^3} h^*_3\left (h^*_ 2\sqrt {E_2^{*2} - 
	m_ 2^2}\left (E^*_2 m_ 3\left (4E^*_2 - m_ 3 \right) - 
m_ 2^2\left (E^*_2 + 2 m_ 3 \right) \right)\right.\\ 
&\left.+ \left (E_2^{*2} - 
m_ 2^2 \right)\left (m_ 3\left (4E^*_2 - 
m_ 3 \right) - 3 m_ 2^2 \right) \right)\,.
\end{split}
\end{equation}

Similarly, in the lab-frame, the differential decay width for the three-body decay is given by
\begin{eqnarray}
   \frac{d^2\Gamma}{dE_2d\cos\theta_2}=\mathcal{F}_1(E_3,E_2,m_3,m_2,h_2,h_3) + \mathcal{F}_2(E_3,E_2,m_3,m_2,h_2,h_3) \cos\theta_2 +  \mathcal{F}_3(E_3,E_2,m_3,m_2,h_2,h_3) \cos\,2\theta_2\,,\nonumber\\
   \label{eq:2dgamma_a}
\end{eqnarray}
where the functions $\mathcal{F}_{1,2,3}$ are given by

\begin{equation}
\begin{split}
\mathcal{F}_1(E_3,E_2,m_3,m_2,h_2,h_3)&= -\frac{G_\nu^2}{384 \pi ^3 E_3} \sqrt{E_2^2-m_2^2}\times
\left[
   \left(h_2 \sqrt{E_2^2-m_2^2}+E_2\right)
           \left(6 E_2 E_3   
              \left(h_3 \sqrt{E_3^2-m_3^2}+E_3\right)\right.\right.\\ &\left.\left. -m_3^2      \left(h_3 \sqrt{E_3{}^2-m_3^2}+2 E_2+3 E_3\right) 
     \right)
      -m_2^2 \left(h_2 \sqrt{E_2^2-m_2^2} \left(h_3 \sqrt{E_3{}^2-m_3^2}+E_3\right) \right.\right.\\ &\left.\left.
              (3 E_2+2 E_3) \left(h_3 \sqrt{E_3{}^2-m_3^2}+E_3\right)-4 m_3^2\right) 
 \right]\,,
\end{split}
\end{equation}
\begin{equation}
\begin{split}
\mathcal{F}_2(E_3,E_2,m_3,m_2,h_2,h_3) &= \frac{G_\nu^2}{384 \pi^3 E_3}\times 
\left[ 
     h_2 \sqrt{E_2^2-m_2^2} 
   \left(   
        h_3 
       \left ( E_2 \left(8 E_2 E_3^2-m_3^2 (4 E_2+E_3)\right) -m_2^2\left(E_3\times\right.\right.\right.\right. \\
      & \left.\left.\left.\left.  (E_2+4 E_3)-2 m_3^2\right)\right) + \sqrt{E_3^2-m_3^2} \left( E_2  \left(8 E_2 E_3-3 m_3^2\right) -m_2^2 (E_2+4 E_3) \right) \right)\right. \\
      &\left. + \left(E_2^2-m_2^2\right) \left(h_3 \left(-m_3^2 (4 E_2+E_3)-3 m_2^2 E_3+8 E_2 E_3^2\right) + \sqrt{E_3^2-m_3^2} \left(-3 m_2^2\right.\right.\right.\\ 
      &\left.\left.\left. -3 m_3^2+8 E_2 E_3 \right) \right)
\right]\,,
\end{split}
\end{equation}
and
\begin{equation}
\begin{split}
\mathcal{F}_3(E_3,E_2,m_3,m_2,h_2,h_3)&=\frac{G_\nu^2}{96 \pi ^3 E_3} \sqrt{E_2^2-m_2^2} \left(E_2 \left(h_2 \sqrt{E_2^2-m_2^2}+E_2\right)-m_2^2\right)\left(m_3^2-E_3\left(h_3 \sqrt{E_3^2-m_3^2}+E_3\right)\right) \,.
\end{split}
\end{equation}

\section*{Acknowledgements}
We thank Kevin Kelly, Yuber Perez-Gonzalez, and Ivan Martinez-Soler for helpful discussions. This work was supported in part by the US Department of Energy (DOE) grant \#de-sc0010143 and in part by the NSF grant PHY-1630782. The work of AdG was also supported in part by the National Science Foundation under Grant No. NSF PHY-1748958.

\bibliographystyle{kpmod}
\bibliography{DecayNu}
\end{document}